\documentstyle[aps,prl,floats,epsf]{revtex}

\def\be{\begin{equation}}
\def\ee{\end{equation}}
\def\bea{\begin{eqnarray}}
\def\eea{\end{eqnarray}}
\def\Tr{\mathop{\rm Tr}}

\begin{document}
\bibliographystyle{prsty}
\title{Weak Charge Quantization on  Superconducting Islands}
\draft
\author{M. V. Feigelman$^1$, A. Kamenev$^{2}$, A. I. Larkin$^{1,2}$
and M. A. Skvortsov$^1$}

\address{$^1$L. D. Landau Institute for Theoretical Physics,
Moscow 117940, Russia}
\address{$^2$Department of Physics, University of Minnesota,
Minneapolis, MN 55455, USA}
\date{\today}

\maketitle

\begin{abstract}
    We consider the Coulomb blockade on a superconductive quantum
    dot strongly coupled to a lead through a tunnelling barrier and/or
    normal diffusive metal. Andreev transport of the correlated
    pairs leads to  quantum fluctuations of the charge on the
    dot. These fluctuations result in exponential
    renormalization of the effective charging energy. We employ
    two complimentary ways to approach the problem, leading to the
    coinciding results: the instanton  and the functional RG treatment
    of the non--linear sigma model. We also derive the charging
    energy renormalization in terms of arbitrary transmission
    matrix of the multi--channel interface.
\end{abstract}


\bigskip

\section{Introduction}

Physics of interacting electronic systems in  presence of disorder
has been a subject of an intense study  already for a few decades
\cite{book}. Various theoretical approaches have been developed
for the description of both metallic and insulator phases. The
non--linear $\sigma$--model (NL$\sigma$M) in the replica
\cite{Finkelstein90,Belitz94} or dynamic (Keldysh)
\cite{Kamenev99,Chamon99} formulation has proven to be the most
powerful tool to deal with  the weakly disordered (metallic)
phase. It was shown that both versions of the $\sigma$--model may
effectively treat the perturbation theory as well as  the
renormalization group (RG) formalism. Unlike the non--interacting
models, where a whole specter of non--perturbative results is
available \cite{Efetov}, it was relatively little progress in the
development of non--perturbative solutions of the interacting
NL$\sigma$M. Our goal is to make a  step in this direction, using
the Coulomb blockade (CB) on a superconductive (SC) quantum dot as
a prototypical example.

The Coulomb blockade  on a quantum dot coupled to a certain number
of leads proved to be an extremely rich and fascinating phenomena
both theoretically and experimentally (see
Refs.~\cite{Likharev,Aleiner} for review). From the theoretical
point of view it provides a model, where the Coulomb interactions,
being spatially localized, may be treated in a non--perturbative
way. The interactions strongly affect charge (and spin)
fluctuations between the dot and the leads, that manifests in the
peculiar transport and thermodynamic behavior of the coupled
dot--lead system. In the case of practically isolated dot, such
that the dimensionless conductance (measured in units of
$G_Q\equiv e^2/(2\pi\hbar)$) of the dot--lead interface is small
($G\ll 1$), the fluctuations usually may be taken into account
perturbatively. (The notable exceptions are provided by the
vicinity of the charge degeneracy point and by the Kondo effect on
the dot \cite{Aleiner}). Here we concentrate on the opposite
scenario of the dot strongly connected to the leads ($G > 1$). In
this case the Coulomb blockade is expected to be suppressed by the
charge fluctuations and the overall effect of interactions on a
single dot to be weak. It is, however, a challenging theoretical
problem, lacking an obvious small parameter, to understand a
remnants of the CB on a strongly connected dot. More importantly
the weak CB may prove to be a strong phenomena in the granulated
systems, with many (superconductive) dots incorporated in the
conducting matrix.

For the normal dot the weak CB effect was relatively well
understood from the various points of view.
Matveev~\cite{Matveev95} gave a complete picture of the phenomena
for the case of one or two conducting channels connecting the dot
and the leads. For the multi--channel system the CB suppression
was calculated with the exponential accuracy employing RG
technique~\cite{Schon,slf01}, the instanton
calculus~\cite{Zaikin,Grabert,Nazarov99,Kamenev00} and
bosonization~\cite{Aleiner}. The factor, suppressing the
thermodynamic CB oscillations, was formulated by
Nazarov~\cite{Nazarov99} in terms of transmission coefficients,
$T_\alpha$, of the dot--lead interface as $\prod_\alpha
(1-T_\alpha)^{1/2}$, where $\alpha = 1\ldots N$ and $N$ is the
number of {\em spinless} channels. The remarkable feature of this
result is that the essentially many--body phenomena may be
described via the single--particle (non--interacting) scattering
matrix only. Yet the knowledge of the interface conductance
$G=\sum_\alpha T_\alpha$ alone is not sufficient to describe the
CB oscillations. For example, in a dot coupled to a lead through
the tunnelling barrier with $T_\alpha \ll 1$, one obtains for the
CB suppression factor $\exp\{-G_T/2\}$. Another important case of
a dot--lead interface is a coupling via a piece of diffusive
metal. Employing Dorokhov statistical distribution~\cite{Dorokhov}
of the transmission coefficients $P(T)= G_D/(2T\sqrt{1-T})$, where
$G_D$ is the conductance of the diffusive area,  one obtains for
the typical CB suppression factor $\exp\{-\pi^2 G_D/8\}$. Since
the result is exponentially sensitive -- the difference between
e.g. $G/2$ and $\pi^2 G/8$ may be actually enormous.

The superconductive  dot in contact with normal leads is even more
challenging system. We  consider the low temperatures  $T\ll
\Delta$, where $\Delta$ is the SC gap, and therefore only very few
(if at all) quasi--particle excitations are allowed to leave or
enter the superconductor. The dominant mechanism of the charge
transfer through the interface is thus the Andreev transport of
the correlated pairs. The Andreev transmission of a given channel
is known to be ${\cal T}_\alpha =
T_\alpha^2/(2-T_\alpha)^2$~\cite{Beenakker92,Lambert93}. For the
tunnelling barrier setup ($T_\alpha \ll 1$) this leads to an
overall Andreev conductance that scales like $G_A \sim G_T^2/N$
(pair tunnelling probability). In most cases this is a very small
number. The presence of the diffusive normal metal adjacent to the
tunnelling barrier increases the Andreev conductance to $G_A\sim
G_T^2/G_D$ (or, in the case $G_T>G_D$, to $G_A\sim G_D$). The
physical reason of this phenomena is multiple attempts of Andreev
transmission due to the coherent back--scattering on the normal
impurities. The natural question is whether it is $G$ or $G_A$ (or
may be none of them) that determines the amplitude of the CB
oscillations. In case $G_A$ is the relevant quantity, one may
wonder whether the coherent back--scattering enhancement should be
taken into account.   The answers are not immediately obvious,
since while the SC strongly prefers the pair transport -- the
Coulomb energy of the dot makes the entrance of two charges at
once energetically costly. The interactions of the dot may also
provide a dephasing mechanism which ruins coherent
back--scattering. We show that parametrically it is indeed $G_A$
(including diffusive enhancement) which determines the CB
amplitude. The coefficient in the exponent, however,  is setup
dependent, making the final answer  sensitive to the ratio
$G_T/G_D$ and not only to $G_A$. The effective charging energy
turns out to be proportional  to $\prod_\alpha (1-{\cal
T}_\alpha)^{1/2}$, where ${\cal T}_\alpha$ are the Andreev
transmissions defined above. This expression provides a remarkable
analogy between the normal and Andreev results.

The other interesting  phenomena brought by superconductivity is
the parity effect \cite{Hekking93,Nazarov94}. At very small
temperature the pair tunnelling is the dominant mechanism and the
period (in the gate voltage) of the CB oscillations is twice
larger than in the normal dot. For a closed dot the parity
phenomena is destroyed by the entropic effects at moderately small
temperature $T^* =\Delta/\ln(\Delta/\delta) \ll \Delta$, where
$\delta$ is the mean single--particle level spacing on the dot
\cite{Nazarov94}. The physical reason for this temperature to be
much less than $\Delta$ is that it is enough to have a single
excited quasiparticle to destroy the parity effect.  At larger
temperature the system exhibits the ``normal'' oscillation period.
We show that for an open dot the transition temperature between
the normal and doubled period is somewhat larger than for a closed
dot and is given by $T^\dagger =\Delta/\ln(\Delta/G_A \delta)$,
provided $G_A\delta < \Delta$.

Technically we treat the problem from the two seemingly distinct
perspectives. First we look for the spatially--dependent instanton
solution of interacting NL$\sigma$M on SC dot in contact with the
normal diffusive region. The finite action of such instanton
configuration results in the exponential suppression of CB
amplitude. Alternatively we can treat the problem employing the
functional RG technique. The latter approach encodes the entire
dot--lead interface in an (infinite) set of coefficients, whose
values are subsequently renormalized by the quantum fluctuations
of the phase on the  dot. Renormalization is terminated at the
cut--off energy scale where the conductance reaches unity. The
renormalized (exponentially small) cut--off energy is the
effective charging energy $\tilde E_C$, this dictates the
amplitude of the CB oscillations. One of the messages of the
present paper is that these two approaches lead to the identical
results.

The paper is organized as follows: in section \ref{s2} we describe
the setup and formulate the appropriate action in the interacting
NL$\sigma$M language. Section \ref{s3} is devoted to the instanton
treatment of the problem in the real space. In section \ref{s4} we
derive the proximity action functional and obtain CB amplitude for
an arbitrary set of transmission eigenvalues (using instanton
approach) at moderately low temperatures $\tilde E_C \ll T \ll
E_C$. Then in section \ref{s5} the RG approach is employed to
treat the CB in the same temperature range. It is demonstrated
that the results coincide with those obtained by instanton
techniques. Finally in section \ref{s7} we discuss the physical
results and their possible experimental signatures. Appendix
contains  derivation of the instanton action starting from the
real-time Keldysh functional technique.

\section{Problem Setup and  Action}
\label{s2}

We consider  a large diffusive (or chaotic) SC   dot. The mean
single--particle level spacing of the dot, $\delta$, is supposed
to be the smallest energy scale in the problem. The SC gap,
$\Delta$, on the other hand, is the largest scale. The electrons
on the dot interact via the capacitive interaction of the form
\begin{equation}
H_{int}=E_C(\hat N -q)^2 \, ,
                                              \label{Hint}
\end{equation}
where $\hat N$ is the electron number operator and $q$ is the
rescaled dimensionless gate voltage potential. The charging
energy, $E_C=e^2/(2C)$, is assumed to satisfy the inequalities
$\delta < E_C < \Delta$.

\begin{figure}
\centerline{\epsfbox{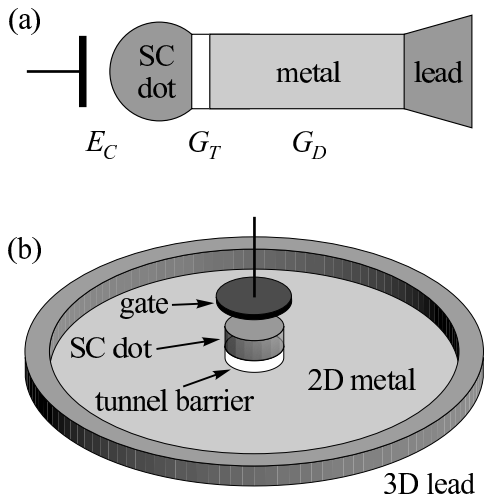}}
\vspace{4mm}
\caption{(a) Schematic view of a SC dot connected to a bulk lead through
a tunnel barrier with the conductance $G_T$ and a piece of diffusive
metal with the conductance $G_D$. The charging energy $E_C$ controls
the coupling between the dot and the gate.
(b) A possible 2D realization.}
\label{fig1}
\end{figure}

The dot is separated from the normal diffusive metal by the
tunnelling barrier with the conductance $G_T$, see Fig.~\ref{fig1}.
The piece of the quasi 1D or 2D metal having  size $L$
and conductance $G_D$ is in turn connected  to a clean bulk 3D
lead. By the reasons which are explained later we shall assume
that the Thouless energy of the diffusive region, $E_{\rm Th} =
D/L^2$, where $D$ is the diffusion coefficient, is larger than the
charging energy of the dot, $E_{\rm Th}>E_C$. The opposite limiting
case requires a separate treatment and will be presented
elsewhere.

We shall be interested in those thermodynamic characteristics of
the dot, which depend in an oscillatory way on the gate voltage
potential, $q$. Specifically,  we look for the free energy
$F(q)=-T\ln Z(q)$, where $Z(q)$ is the partition function. Both of
these quantities depend on a particular realization of disorder in
the diffusive region and inside the dot. We shall therefore look
for the disorder averaged quantity like $\langle F(q) \rangle$.
For an open dot $F(q)$ is an oscillatory function, with a phase
sensitive to a disorder realization.  It is therefore convenient
sometimes to calculate also the correlation function $\langle F(q)
F(q') \rangle$ \cite{Kamenev00}. The latter carries information
about the typical $F(q)$, rather than the average one (that may
spuriously vanish due to phase randomness).

The disorder averaging may be performed in two ways either by
introducing  $p$ replica of the system
\cite{Finkelstein90,Belitz94} and sending $p\to 0$ in the end, or
by dealing with the dynamical Keldysh formulation
\cite{Kamenev99,Chamon99}. We shall employ both of these
approaches to demonstrates what they are consistent and
interchangeable.  In the replica formalism, the NL$\sigma$M is
formulated in terms of the matrix field theory of the matrix field
$Q^{ab}_{ij}({\bf
 r},\tau,\tau')$, where $a,b=1,2\ldots 2p$ are the replica indexes
(one needs $2p$ replica to describe the correlation function
$\langle Z^p(q) Z^p(q') \rangle$) and $i,j =1,2$ are Gorkov--Nambu
indexes. The correlation function of the free energies is given by
$\langle F(q) F(q') \rangle = \lim_{p\to 0} p^{-2}(\langle Z^p(q)
Z^p(q') \rangle - 1)$.  The $Q$--matrix obeys the constraint $Q^2
= 1$ and the fermionic anti--periodic boundary condition in both
of its imaginary time arguments $\tau,\tau' \in [0,\beta]$.  The
coordinate ${\bf r}$ runs over the volume of the SC dot and the
normal diffusive region. The matrix $Q$--field describes dynamics
of the electrons; it is coupled to a scalar bosonic vector field
$\Phi_a(\tau)$ which originates from the Coulomb interactions on
the dot. Since we restrict ourselves by the simplest capacitive
interaction Eq.~(\ref{Hint}), the field $\Phi_a(\tau)$ is space
independent throughout the dot  and vanishes outside the dot. As a
result, only spatially independent (${\bf q} =0$) component of the
superconductive $Q$--matrix of the dot appears to be coupled to
the Coulomb field $\Phi$. Therefore we assume that the $Q$--matrix
is spatially constant inside the dot, while it may have a
non--trivial spatial structure inside the normal diffusive region.
The effective action for our geometry contains three terms
\begin{equation}
S = S_{dot} + S_T + S_{D}\, ,
                                            \label{action}
\end{equation}
where $S_{dot}$ and $S_{D}$ are the effective actions of the
isolated SC dot and normal diffusive region correspondingly, while
$S_T$ is the tunnelling action which provides the  coupling
between the two. We shall examine separately these three
contributions.

\subsection{Action of the dot}

According to our model -- the dot is the only region, where
electrons interact via the Coulomb interaction Eq.~(\ref{Hint}).
As a result, the dot's action contains the two coupled fields:
$\Phi_a(\tau)$ and $Q_S^{ab}(\tau,\tau')$. The first one
originates from the Hubbard--Stratonovich decoupling of the
Coulomb term \cite{Kamenev96}, while the latter is the result of
disorder averaging and integration of the fermionic degrees of
freedom on the dot \cite{Finkelstein90}. Both of these fields are
spatially independent, reflecting the fact that the Thouless
energy of the dot is assumed to be large. The action takes the
standard by now form:
\begin{equation}
S_{dot}[Q_S,\Phi] =  \sum\limits_{a=1}^{2p} \int\limits_0^\beta
\!\! d\tau \left( {\dot\Phi_a^2 \over 4 E_c} - i q^a \dot\Phi_a
\right) - {\pi\over \delta} \, \Tr \left\{
\left(\sigma_3\partial_\tau - i\dot\Phi + \hat\Delta \right) Q_S
\right\} \, .
                                            \label{dot}
\end{equation}
Here the replica vector of gate voltages, $q^a$, is defined in a
way that $q^a=q$ for $a\in[1,p]$ while $q^a=q'$ for $a\in
[p+1,2p]$. The SC order parameter $\hat\Delta=(\Delta_a\sigma_+ -
\Delta_a^{*}\sigma_-)\delta^{ab}$, is the same in all replica and
is written as a matrix in the Gorkov--Nambu space, where
$\sigma_{\pm}=(\sigma_1\pm i\sigma_2)/2$. Notice, that the gate
voltage dependence of the partition function originates entirely
from $\exp\{ i \pi \sum_a q^a W_a \}$, where $\pi W_a \equiv
\int\! d\tau\, \dot \Phi_a(\tau) = \Phi_a(\beta) - \Phi_a(0) $ is
the zeroth Matsubara component of the $\dot \Phi_a(\tau)$ field.

Since all the energy scales we consider are larger than $\delta$,
one may evaluate the second term in the dot's action~(\ref{dot})
in the saddle point approximation over the field $Q_S$. In so
doing, one disregards the mesoscopic conductance fluctuations of
the dot--lead interface \cite{Aleiner}. The saddle point value of
the $Q_S$ field is given by the Gorkov Green function gauged by
the phase $\Phi(\tau)$:
\begin{equation}
Q_S^{ab}(\tau,\tau') = e^{i\sigma_3\Phi_a(\tau)}
\Lambda^{ab}_S(\tau,\tau') e^{-i\sigma_3\Phi_b(\tau')}\, .
                                                      \label{Qdot}
\end{equation}
The Gorkov Green function, $\Lambda_S$, has the standard form,
which is more familiar in the Matsubara basis (we assume the phase
of the SC dot without the Coulomb interactions to be zero)
\begin{equation}
\Lambda^{ab}_S(n,m)=\delta^{ab} \delta_{nm}\left(\begin{array}{cc}
\cos \theta_n & \sin \theta_n \\
\sin \theta_n  & -\cos \theta_n
\end{array} \right)\, ,
                                                  \label{lambda}
\end{equation}
where
\begin{equation}
\cos \theta_n \equiv \frac{\epsilon_n}{\sqrt{\epsilon_n^2
+|\Delta|^2}}\,; \hskip 1cm \sin \theta_n \equiv
\frac{|\Delta|}{\sqrt{\epsilon_n^2 +|\Delta|^2}}\,
                                                  \label{theta}
\end{equation}
and $\epsilon_n=\pi T(2n  + 1)$ is the fermionic Matsubara
frequency. The phase rotation~(\ref{Qdot}) preserves the fermionic
anti--periodic boundary conditions if all $W_a$ are {\em even}
integers. At small temperatures, $T\ll \Delta$, however, one has
$\cos \theta_n \approx 0$ and $\sin \theta_n \approx 1$ and
therefore the Gorkov matrix, $\Lambda_S$ is (almost) off--diagonal
in the Nambu space and local in time. As a result, only
$2\Phi(\tau)$ participate in the phase rotation, Eq.~(\ref{Qdot}).
Therefore the {\em odd}--integer $W_a$ preserve the fermionic
boundary conditions, as well. The odd--integers $W_a$ result in
the doubling of the period of the $F(q)$ function with respect to
the normal case. This reflects the fact that the pair transfer is
the dominant mechanism of the charge exchange.

In fact, one has to be more careful and recall that, according to
Eq.~(\ref{dot}), one has to perform integrations over all
Matsubara components of $\dot\Phi_a(\tau)$ fields
\cite{Kamenev96}. All non--zero Matsubara components may be
eliminated by the gauge transformation, Eq.~(\ref{Qdot}). As a
result, they have no effect on the thermodynamic of an {\em
isolated} dot at all (they are of major importance, of course,
once the dot is coupled to the leads). The remaining (usual)
integral over the zeroth Matsubara component, $W_a$, must be
performed explicitly. To this end one notices that $\pi T W_a$
enters the action as an imaginary chemical potential in the
replica $a$. The (replicated) free energy of an isolated dot is
thus a periodic function of each of $W_a$ with the  period $2$
(indeed the chemical potential always enters as $\exp\{\mu \hat
N/T\}$, since the number operator, $\hat N$, has only integer
eigenvalues -- the periodicity is apparent). The free energy
possesses deep minima at even integer values of $W_a$ with the
quadratic behavior in their vicinity $F_{dot}(W)\approx \pi^2 T^2
W^2/(2\delta)$, where $\delta= (\partial^2 F_{dot}/\partial
\mu^2)^{-1}$ is the mean level spacing of the dot. For a
sufficiently large dot, where $\delta<T$, the integrals over $W_a$
may be performed in the saddle point approximation, which results
in the even--integer quantization of $W_a$. In a SC dot there are
additional minima at odd--integer values of $W_a$. Consequently
the integration over $W_a$ is substituted by the summation over
all integers. At $T>0$ the additional minima of the free energy at
odd integer $W_a$ are not as deep as at the even integers,
reflecting the fact that the addition of an odd number of
electrons is possible by creating a quasi--particle. We shall
evaluate now the action cost of the odd minima with respect to the
even ones.

For the even values, say $W_a=0$, one substitute the saddle point
solution, Eq.~(\ref{Qdot}), into the action, Eq.~(\ref{dot}), and
obtains (we disregard for a moment the first term in
Eq.~(\ref{dot}))
\begin{equation}
S_{dot}(W_a=0) = -{\pi\over \delta}\, \Tr\{ (\sigma_3
\partial_\tau + \hat \Delta)\Lambda_S^{aa}\} = -{2\pi\over
\delta}\sum\limits_n \sqrt{\epsilon_n^2 +|\Delta|^2}\, .
                                                  \label{integer}
\end{equation}
This sum is divergent. However, it is only the difference of the
action between different replica that has a physical significance.
The later quantity is convergent as we shall see momentarily. For
the odd integers, say $W_b=1$, there is an imaginary component
$i\pi T$ of the chemical potential in Eq.~(\ref{dot}). It may be
eliminated by the gauge transformation which converts the
antiperiodic boundary conditions for the fermions into the
periodic one. As a result, one arrives to the same expression as
Eq.~(\ref{integer}) with the  fermionic Matsubara frequency
$\epsilon_n=\pi T(2n +1)$  substituted by the bosonic one
$\omega_n=2\pi T n$:
\begin{equation}
S_{dot}\left( W_b=1 \right) = -{2\pi\over \delta}\sum\limits_n
\sqrt{\omega_n^2 +|\Delta|^2}\, .
                                                  \label{half}
\end{equation}
Employing the Poisson summation formula, one  obtains for the
difference
\begin{equation}
{\cal A}(T) \equiv S_{dot}(1)-S_{dot}(0) = {8|\Delta|\over \delta}
\sum\limits_{l=0}^{\infty} {1\over 2l+1}
K_1\left((2l+1){|\Delta|\over T} \right) \approx  {4\sqrt{2\pi
|\Delta| T} \over \delta}\,  e^{-|\Delta|/ T}\, ,
                                                  \label{diff}
\end{equation}
where the last equality assumes $T\ll \Delta$. In the opposite
limit of  the normal dot, $T > \Delta$, one finds ${\cal
A}(T)\approx \pi^2 T/\delta$, reflecting the fact that in a normal
dot the odd--integer minima are absent. Although these minima
persist up to $T\approx \Delta$, their   contribution is
exponentially suppressed at $T >T^*$, where $T^*$ is determined
from the condition $S_{dot}(1)-S_{dot}(0)\approx 1$
\cite{Hekking93,Nazarov94}
\begin{equation}
T^* \approx \frac{|\Delta|}{\ln |\Delta|/\delta } \ll |\Delta|\, .
                                                  \label{Tstar}
\end{equation}
As a result, there is an important temperature dependence
associated with the odd--integer winding numbers at the scale
$T\approx T^* \ll |\Delta|$. We shall see below that for the dot
strongly coupled to the leads, the corresponding temperature scale
is slightly different.

We summarize now our discussion of the action of a large SC dot
with the Coulomb interaction at $T\ll \Delta$. The scalar
potential in each replica obeys the boundary condition
\begin{equation}
\Phi_a(\beta) - \Phi_a(0) = \pi W_a\, ,
                                                  \label{winding}
\end{equation}
where $W_a$ is an  integer winding number. The corresponding
saddle point value of the $Q_S$--matrix field is given by
Eq.~(\ref{Qdot}). The action of the dot takes the form:
\begin{equation}
S_{dot}[\Phi] =  \sum\limits_{a=1}^{2p} \left[ -i \pi q^a W_a +
{\cal A}(T)
\delta_{W_a \bmod 2,1} +  \int\limits_0^\beta \!\! d\tau\,
{\dot\Phi_a^2  \over 4 E_C} \right] \, ,
                                            \label{dot1}
\end{equation}
where ${\cal A}(T)$ is given by Eq.~(\ref{diff}).

\subsection{Tunnelling barrier and diffusive region action }
\label{s23}

The tunnelling action couples the $Q_S$--field  on the dot with
the $Q({\bf r}=0)$--field at the point ${\bf r}=0$  adjacent    to
the tunnelling barrier from  the normal metal  side. It has the
standard form \cite{Efetov}
\begin{equation}
S_T = - {G_T \over 8}\,  \Tr\{Q_S Q(0)\}\, ,
                                     \label{ST}
\end{equation}
where $G_T$ is the tunnelling conductance.

The action of the normal diffusive region also has  the standard
form \cite{Finkelstein90,Efetov}
\begin{equation}
S_D = \frac{\pi\nu}{4}  \int\limits_0^L\!\!\! d{\bf r}\,
D\Tr\{(\nabla Q({\bf r}))^2\}\, ,
                                     \label{SD}
\end{equation}
where $D$ is the diffusion constant and $\nu$ the density of
states (per spin) of the normal region. The total conductance of
the normal region is given by $G_D =4\pi\nu D/L$ for the quasi 1D
geometry of the normal region and $G_D=8\pi^2 \nu D/\ln (L/d)$ for
the 2D geometry. Here $L$ is the length (radius) of the 1D (2D)
region and $d$ is the radius of the SC dot. We have omitted the
frequency term $\nu \Tr\{\epsilon Q\}$ on the r.h.s. of
Eq.~(\ref{SD}), because of the assumption that $E_{\rm Th}>E_C>T$.  At
the point where the diffusive region is attached to the normal
bulk lead one has to impose the boundary condition
\begin{equation}
Q({\bf r}=L) =  \Lambda_N\, ,
                                     \label{bc}
\end{equation}
where
$\Lambda^{ab}_N(n,m)=\sigma_3\delta^{ab}\delta_{nm}\mbox{sign}(\epsilon_n)$
is the appropriate $Q$--matrix of the normal bulk lead.

Alternatively, one may imagine integrating out the $Q({\bf r})
$--field of the normal region, subject to the boundary condition
Eq.~(\ref{bc}) and weighted by the action $S_T+S_D$. This
procedure (we shall describe it in details in section \ref{s4})
leads to the effective action $S_{TD}$ of the interface plus
diffusive region written in terms of $Q_S$ and $\Lambda_N$ only.
If all the relevant energy scales are less than the Thouless
energy of the diffusive metal -- the general form such action may
take is \cite{slf01}
\begin{equation}
S_{TD} =
-{G_D\over 8}\, \sum_{l=1}^{\infty}
\gamma_l\, \Tr\{(Q_S \Lambda_N)^l\}\, ,
                                     \label{STSD}
\end{equation}
where $\gamma_l$ are  coefficients which depend on the details of
the interface (in our case the ratio $G_T/G_D$). The largeness of
the Thouless energy is necessary to disregard the retardation
effects and thus possible time non--local coupling between $Q_S$
and $\Lambda_N$. Under such condition, the proximity action,
Eq.~(\ref{STSD}), is completely equivalent to those given by
Eqs.~(\ref{ST}) and (\ref{SD}) upon the proper choice of the set,
$\gamma_l$ \cite{slf01}.

\section{The Instanton Approach}
\label{s3}

We are interested  in the limit of strong coupling between the dot
and the leads, meaning $G_D, G_T > 1$ (the weak coupling limit may
be treated in the spirit of Refs.~\cite{Glazman90,Matveev91,Glazman98}).
For $G_D\gg 1$ the fluctuations of the $Q$--field around its optimal
value are suppressed. One may employ therefore the stationary
phase treatment of the NL$\sigma$M for the dot--lead interface
\cite{Kamenev00}. Taking the variation of the action
Eqs.~(\ref{ST}), (\ref{SD}) under the condition $Q^2 =1$ one
obtains the Usadel equation
\begin{equation}
2\pi\nu \nabla(D Q\nabla Q) - \delta({\bf r}) G_T [Q,Q_S] = 0\, .
                                           \label{Usadel}
\end{equation}
This equation is to be solved for a fixed $Q_S = Q_S[\Phi]$ given
by Eq.~(\ref{Qdot}) and with the boundary conditions
Eq.~(\ref{bc}). The solution $Q=Q[\Phi]$ after being substituted
back into the action Eqs.~(\ref{ST}), (\ref{SD}) -- results in the
semiclassical phase action $S[\Phi]$. The later may then be
investigated using the instanton approach applied to the
$\Phi(\tau)$--field.

\subsection{Zero winding number}
\label{s31}

As a warm--up exercise, consider the zero winding number sector of
the theory, $W_a = 0$ for $a=1,\ldots 2p$. Obviously it does not
produce an oscillatory dependence of $F(q)$, cf.\ Eq.~(\ref{dot1}),
and therefore serves only an axillary purpose. The lowest energy
configuration in zero winding number sector is $\Phi_a(\tau)=0$
and therefore $Q_S$ on the dot is simply given by the BCS
$\Lambda_S$, Eq.~(\ref{lambda}). Solution of the Usadel equation
(\ref{Usadel}) may be written as
\begin{equation}
Q({\bf r}) = \Lambda_N \exp\left\{ i\,u({\bf r})\, \theta
\otimes\sigma_2 \right\} \, ,
                                                  \label{zerosector}
\end{equation}
where $u({\bf r})$ is the normalized ``voltage drop'' inside the
normal region: $u(r) = (L- r)/L\,$  in 1D and $u(r) = \ln
(L/r)/\ln (L/d) $ in 2D. The absolute value of the voltage drop,
$\theta$, is coordinate independent diagonal (in replica and
Matsubara space) matrix. It has a physical meaning of the SC
rotation angle of the normal metal in the direct proximity to the
dot. One may substitute the solution back into the action to
obtain (for a single replica and Matsubara component)
\begin{equation}
S_0= {1\over 8} \left[ G_D \theta^2 - 2
G_T\cos(\theta_n-\theta)\right]\, ,
                                                \label{zeroS}
\end{equation}
where the subscript ``$0$'' stresses that we work with zero
winding numbers and $\theta_n$ is defined by Eq.~(\ref{theta}).
For small energy, $\epsilon_n\ll \Delta$, one has $\theta_n\approx
\pi/2$ and therefore the corresponding action takes the form
\begin{equation}
S_0 = {G_D\over 8} \left[\theta^2 - 2 t \sin \theta\right]\, ,
                                                \label{zeroS1}
\end{equation}
where $t\equiv G_T/G_D$. This action is minimized when $\theta =
\theta(t)$ satisfies the equation
\begin{equation}
\theta = t \cos \theta\, .
                                                \label{zeroUsadel1}
\end{equation}
The lowest energy solution of this  equation smoothly interpolates
between $\theta=0$ for $t\ll 1$ and $\theta=\pi/2$ for $t\gg 1$.
Finally, this solution has to be substituted into
Eq.~(\ref{zeroS1}) to find the action cost, $S_0=S_0(t)$, for the
zero winding number  configuration.

\subsection{Non--zero winding numbers}
\label{s32}

To calculate the oscillatory component of the free energy, $F(q)$,
one has  to consider the $\Phi$--field configurations with
non--zero winding numbers, cf.\ Eq.~(\ref{dot1}). Consider, thus,
the simplest even configuration of winding numbers with $W_1=2$
and all others $W_a=0$ in the remaining $2p-1$ replica. With the
exponential accuracy it is sufficient to consider the ``straight''
windings: $\Phi_{1}(\tau) =  2\pi T \tau$. The saddle point of the
SC $Q$--field on the dot is given by Eq.~(\ref{Qdot}) and takes
the form
\begin{equation}
Q_S^{11}(n,m) =\left( \begin{array}{ll} \delta_{n,m}\cos
\theta_{m-1} & \delta_{n,m+2} \sin\theta_{m+1}  \\
\delta_{n,m-2} \sin\theta_{m-1}  & - \delta_{n,m}\cos \theta_{m+1}
\end{array} \right)\, ,
                                           \label{W1}
\end{equation}
where the $2\times 2$ structure  is the Nambu space. In all other
replica except of $a=1$ the $Q_S$--matrix has the form
Eq.~(\ref{lambda}). One may check that the Usadel equation is
solved by exactly the same $O(2)$ rotation as in $W=0$ case,
Eq.~(\ref{zerosector}), performed in each of the following
$2\times 2$ Nambu blocks
\begin{equation}
\left( \begin{array}{ll} Q_{11}(n+1,n+1) & Q_{12}(n+1,n-1)  \\
 Q_{21}(n-1,n+1) & Q_{22}(n-1,n-1)
\end{array} \right) =\left( \begin{array}{cc} \cos
\theta_{n} &  \sin\theta_{n}  \\
 \sin\theta_{n}  & - \cos \theta_{n}
\end{array} \right)\, ;
                                           \label{W11}
\end{equation}
the Nambu indexes are explicitly stated on the l.h.s.
 In this $2\times 2$ block $\Lambda_N=\sigma_3$ and therefore the
matrix from the r.h.s of Eq.~(\ref{W11}) may indeed be rotated
into $\Lambda_N$ as in  Eq.~(\ref{zerosector}).  The only
exception are the two lowest Matsubara components $n=0$ and $n=1$,
which form $4\times 4$ block of the form
\begin{equation}
\left( \begin{array}{llll} Q_{11}(1,1) & &Q_{12}(1,-1) & \\
& Q_{11}(2,2) & & Q_{12}(2,0)\\
Q_{21}(-1,1) & &Q_{22}(-1,-1) & \\
& Q_{21}(0,2) & & Q_{22}(0,0)
\end{array} \right) = \left( \begin{array}{cccc}
\cos \theta_{0} & & \sin\theta_{0} &  \\
  & \cos \theta_1  &  & \sin\theta_1 \\
\sin\theta_0 & & -\cos\theta_0 &\\
  & \sin\theta_1  &   &  -\cos\theta_1
\end{array} \right)\, .
                                           \label{44S}
\end{equation}
The corresponding $4\times 4 $ block of the $\Lambda_N$--matrix on
the normal lead  is the unit matrix. Obviously the unit matrix
cannot be unitary rotated  into the block Eq.~(\ref{44S}) and
therefore solution of the Usadel equation in this block is
impossible. The difficulty originates from the fact that, due to
the random phase of the CB oscillations, the average free energy,
$\langle F(q) \rangle$, is not an oscillatory function. We need
therefore to consider winding number configuration of the form
$W_1 =2$ and $W_{p+1}=-2$, while all other $W_a=0$. The
contribution to the correlation function from such a configuration
is proportional to $\exp\{2\pi i (q-q')\}$, that is the lowest
normal  harmonic  of the correlation function $\langle F(q) F(q')
\rangle$. The ``dangerous'' $4\times 4$ block in the $p+1$--st
replica is given by
\begin{equation}
\left( \begin{array}{llll} Q_{11}(-1,-1) & &Q_{12}(-1,1) & \\
& Q_{11}(0,0) & & Q_{12}(0,2)\\
Q_{21}(1,-1) & &Q_{22}(1,1) & \\
& Q_{21}(2,0) & & Q_{22}(2,2)
\end{array} \right) = \left( \begin{array}{cccc}
\cos \theta_{0} & & \sin\theta_{0} &  \\
  & \cos \theta_1  &  & \sin\theta_1 \\
\sin\theta_0 & & -\cos\theta_0 &\\
  & \sin\theta_1  &   &  -\cos\theta_1
\end{array} \right)\, ,
                                           \label{44S1}
\end{equation}
while the corresponding $4\times 4 $ block of the
$\Lambda_N$--matrix is minus one times the unit matrix. Here as
well the unitary rotation between the points ${\bf r}=L$ and ${\bf
r}=0$ is impossible.  However, if one combines $a=1$ and $a=p+1$
replica and allows rotation between them -- then the unitary
rotation may be found \cite{Kamenev00}. Indeed, combining both
``dangerous'' blocks into the single $8\times 8$ block (i.e.
combine Eq.~(\ref{44S}) with Eq.~(\ref{44S1}) on the dot and unit
matrix with the minus unit matrix on the normal lead), one readily
see that they may be unitary connected (since they possess  the
same set of eigenvalues).

The calculations are simplified in the low--temperature case,
$T\ll \Delta$, where $\theta_{n}\approx \pi/2$. In this case the
$8\times 8$ block takes the form
\begin{equation}
Q_S = \left(\begin{array}{ll} 1&0\\0&1 \end{array} \right)
\otimes\left(
\begin{array}{ll}
\sigma_1& 0  \\
0 & \sigma_1
\end{array} \right)\,
                                           \label{88S}
\end{equation}
on the dot and
\begin{equation}
\Lambda_N = \left(\begin{array}{cc} 1&0\\0&1 \end{array} \right)
\otimes \left( \begin{array}{cc}
\sigma_0& 0  \\
0 & -\sigma_0
\end{array} \right)\,
                                           \label{88N}
\end{equation}
on the normal lead ($\sigma_0$ is the unit matrix in the Nambu
space). Here the outer $2\times 2$ unit matrix represents the
space of $n=0$ and $n=1$ Matsubara components, while the inner one
-- the replica space of $a=1$ and $a=p+1$ (Pauli matrices act in
the Nambu space). We seek thus for the solution of the Usadel
equation in the $8\times 8$ sub--space, having in mind that the
solution for all other Matsubara components and replica is exactly
the same as in $W=0$ sector,
Eqs.~(\ref{zerosector})--(\ref{zeroUsadel1}). The general solution
for ${\bf r}\neq 0$ may be written as
\begin{equation}
Q({\bf r}) = \Lambda_N   \exp\left\{ i u({\bf r})\,
\left(\begin{array}{ll} 0&B\\B^\dagger &0\end{array} \right)
\right\}\, ,
                                           \label{solution}
\end{equation}
where $u({\bf r})$ function was defined after
Eq.~(\ref{zerosector}) and $B$ is a coordinate--independent Nambu
matrix. Employing the singular value decomposition, it may be
written as
\begin{equation}
B = U^{-1} \left(\begin{array}{ll} \Theta_1&0\\0& \Theta_2
\end{array} \right) V\, ,
                                           \label{B}
\end{equation}
where $U=e^{i u_3 \sigma_3}e^{i u_2 \sigma_2}e^{i u_1 \sigma_1}$
and $V=e^{i v_3 \sigma_3}e^{i v_2 \sigma_2}e^{i v_1 \sigma_1}$ are
$SU(2)$ matrices and $\Theta_{1}\leq \Theta_2 $ are real singular
values. Substituting the solution back into the action, one
obtains (for each of the two involved Matsubara frequencies)
\begin{equation}
2 S_{\pm 2} = {1\over 4} \left[ G_D(\Theta_1^2 + \Theta_2^2) -
G_T(\sin 2 u_2 - \sin 2 v_2)(\cos\Theta_1 - \cos\Theta_2)\right]
\, ,
                                           \label{SW1}
\end{equation}
where the subscript $\pm 2$ stays for the corresponding winding
numbers and the coefficient two on the l.h.s. reminds that two
replica were involved. The next step is to minimize the action
over $\Theta_{1,2}$, $u_2$ and $v_2$ angles. Three of the four
equations for the minima have only trivial (parameter independent)
solutions: $\Theta_1=0$; $u_2=-\pi/4$ and $v_2=\pi/4$. The action
in terms of the single non--trivial angle $\Theta_{2}$ finally
takes the form
\begin{equation}
2 S_{\pm 2} =  {G_D\over 4} \left[\Theta_2^2  + 2 t (\cos\Theta_2
- 1) \right]  \, ,
                                           \label{SW2}
\end{equation}
where, as above, $t\equiv G_T/G_D$. The corresponding saddle point
equation for $\Theta_{2}=\Theta_2(t)$ is
\begin{equation}
\Theta_2 = t \sin \Theta_2 \, .
                                           \label{Theta2}
\end{equation}
For $t\leq 1$ the only solution of this equation is $\Theta_2=0$,
while for $t\geq 1$ the angle $\Theta_2(t)$ interpolates between
zero (for $t=1$) and $\pi$ (for $t\gg 1$).

\subsection{CB suppression}
\label{s33}

We are now on the position to discuss the suppression of the CB.
Consider first the component of the correlation function $\langle
Z^p(q) Z^p(q') \rangle$ which is proportional to $\cos 2\pi
(q-q')$. As was explained above the relevant field configurations
are those having a single replica with $W_a =\pm 2$, where $a\in
[1,p]$ and a single replica with $W_a =\mp 2$, where $a\in [p+1,
2p]$. The corresponding action is given by $4 {\cal S} \equiv 2
S_{\pm 2} + (2p-2)2 S_0$; (here $S_0$ is multiplied by the number
of replica with zero winding number -- $2p-2$, and by factor of
two, because two Matsubara components are different between $W=\pm
2$ and $W= 0$). Taking the replica limit $p\to 0$ and employing
Eqs.~(\ref{zeroS1}), (\ref{SW2}), one finds
\begin{equation}
2{\cal S}(t) = S_{\pm 2}-2 S_0 = {G_D\over 8}\left[\Theta_2^2(t) -
2\theta^2(t)  + 2t (\cos \Theta_2(t) + 2 \sin \theta(t) -
1)\right] \, ,
                                                \label{CBsupr}
\end{equation}
where $\theta(t)$ and $\Theta_2(t)$ are the solutions of
Eqs.~(\ref{zeroUsadel1}) and (\ref{Theta2}) correspondingly. As a
result, the contribution to the correlation function with the unit
period has the form: $\exp\{-4{\cal S}\} \cos 2\pi (q-q')$, where
the factor of $4$ stays for the fact that two replica are involved
in the correlation function and the winding number is two. We
shall discuss this result in more details in section \ref{s7},
after deriving it using other methods. For the later
reference we need to calculate $t^2\partial_t(2{\cal S}/t)$;
employing the saddle point equations (\ref{zeroUsadel1}) and
(\ref{Theta2}), we obtain
\begin{equation}
t^2\frac{\partial}{\partial t} \left(\frac{2{\cal S}}{t}\right) =
- {G_D\over 8}\left[\Theta_2^2 - 2\theta^2 \right] \, .
                                                \label{t2Sovert}
\end{equation}

As was explained in section \ref{s2}, the parity effects sets in
at small temperature $T < T^*\ll \Delta$. Technically it manifests
itself in the appearance of the odd winding numbers. The
calculations for the field configuration with $W_1 =1$, and
$W_{p+1}=-1$ are exactly parallel to the one for the even winding
numbers. The only difference is that there is the single
``dangerous'' Matsubara frequency, $n=0$, consequently there is no
outer $2\times 2$ Matsubara structure as in Eqs.~(\ref{88S}) and
(\ref{88N}) (the replica and Nambu structures are exactly the
same). As a result, the  action is exactly twice smaller than for
the oscillations with the unit period. We thus obtain for the
component of the correlation function with doubled gate voltage
period: $\exp\{-2({\cal S} + {\cal A}(T)) \}\cos \pi (q-q')$,
where the factor of two in the exponent originates from the two
replica involved.

Finally, let us discuss the technicalities of the replica limit,
$p\to 0$ \cite{Kamenev00}. There are altogether $p^2$ distinct
configurations of the winding numbers which contribute to the
correlation function $\langle Z^p(q) Z^p(q') \rangle$ with the
same oscillatory factor, say $\exp\{2\pi i(q-q')\}$. They
correspond to the  $p$ possible choices for $W_a=2$ and $p$
independent possible choices of $W_a=-2$. There is thus $p^2$
combinatorial factor in the $\langle Z^p(q) Z^p(q') \rangle$ that
is cancelled against the same factor in the denominator for
$\langle F(q) F(q') \rangle = \lim_{p\to 0} p^{-2}(\langle Z^p(q)
Z^p(q')-1) \rangle$.

\section{Proximity action}
\label{s4}

In this Section we discuss the derivation and simple applications
of the proximity action approach to treat the interface between
the SC dot and normal lead. This method  was  recently
suggested~\cite{slf01,slf01short} for the analysis of
superconductive-normal metal transition in a 2D proximity-coupled
array~\cite{fl98,fls01}. Notice that our definition of the running
parameter $\zeta$ and charges $\{\gamma_l\}$ differ from those
used in Refs.~\cite{slf01,slf01short,fls01}, see Appendix
\ref{app1} for details.

\subsection{Derivation of the proximity action}
\label{s41}

Our immediate goal is to derive the proximity action,
Eq.~(\ref{STSD}), starting from Eqs.~(\ref{ST}) and (\ref{SD}). To
this end, let us imagine adding the diffusive metal step by step
with infinitesimal shells having conductance $\delta G \gg G_D>
1$. At some intermediate step of this procedure one has the
tunneling barrier with the part of the normal metal incorporated
into the set of coefficients $\gamma_l(\zeta)$. Here
$\zeta\equiv G_D/G_D(\zeta) \in [0,1]$ is the running parameter
such that $\gamma_l(0) = \delta_{l,1}G_T/G_D$, and $S(0)=S_T$ is the bare
tunneling barrier action; while $\gamma_l(1)=\gamma_l$ and $S(1) = S_{TD}$
is the full proximity action of the barrier ``dressed''  with
the diffusive metal. At each step one adds another thin shell of
the diffusive metal. The action of the entire system takes the
form
\begin{equation}
S(\zeta + \delta\zeta)  = -{G_D\over 8}\, \sum_{l=1}^{\infty}
\gamma_l(\zeta) \Tr\{(Q_S Q)^l\} +  {\delta G\over 16}\, \Tr \{
(Q-\Lambda_N)^2\} \, ,
                                     \label{Szeta}
\end{equation}
where $Q$ is the field right on the boundary between the new shell
and the already  integrated region. The action of the newly  added
shell has the  simple form, $(\delta G/16) \Tr(Q-\Lambda_N)^2$,
since $\delta G\gg G_D > 1$ and therefore the matrix $Q$ has to be
rather close to $\Lambda_N$ (in the leading order in $(\delta
G)^{-1}$ one may disregard all higher order terms
$(Q-\Lambda_N)^l$ with $l > 2$). The next step is to integrate out
the $Q$ field and obtain the new set
$\gamma_l(\zeta+\delta\zeta)$. We perform this procedure in the
saddle point approximation. Taking variation of the action,
Eq.~(\ref{Szeta}), under the condition $Q^2 = 1$ one finds the
saddle point (Usadel) equation
\begin{equation}
G_D \sum_{l=1}^{\infty} l \gamma_l(\zeta) \left[ (Q_S Q)^l - (Q
Q_S)^l \right] - \delta G \left[Q \Lambda_N - \Lambda_N Q\right] =
0 \, .
                                     \label{variance}
\end{equation}
In the leading order in $G_D/\delta G \ll 1$ one obtains for the
saddle point solution
\begin{equation}
Q= \Lambda_N + \frac{G_D}{2 \delta G} \sum\limits_{l=1}^{\infty}
l\gamma_l \left[ (Q_S \Lambda_N)^l \Lambda_N - \Lambda_N (Q_S
\Lambda_N)^l \right]  \, .
                                             \label{spsolution}
\end{equation}
Putting it back into the action, one finally finds
\begin{equation}
S(\zeta+\delta \zeta)= -{G_D\over 8} \left[
\sum\limits_{l=1}^{\infty} \gamma_l \Tr\left\{ (Q_S\Lambda_N)^l
\right\}   + {\delta\zeta \over 2}  \,
\sum\limits_{l,k=1}^{\infty} l k \gamma_l \gamma_k
\Tr\left\{(Q_S\Lambda_N)^{l+k} - (Q_S\Lambda_N)^{|l-k|} \right\}
\right]\, ,
                                           \label{newaction}
\end{equation}
where we employed the fact that $\delta G = G_D/\delta \zeta$. One
ends up with the following evolution equations for
$\gamma_l(\zeta)$:
\begin{equation}
\frac{d \gamma_l}{ d \zeta} = {1\over 2} \sum\limits_{k=1}^\infty
k(l+k) \gamma_k\gamma_{l+k}  -{1\over 4} \sum\limits_{k=1}^{l-1}
k(l-k) \gamma_k\gamma_{l-k}  \, .
                                   \label{evolution}
\end{equation}

To solve this set of equations it is convenient to define the
function $u(\zeta,\Theta) \equiv \sum\limits_{l=1}^{\infty} l
\gamma_l(\zeta) \sin (l\Theta)$. With it's help
Eq.~(\ref{evolution}) takes the form
\begin{equation}
 u_\zeta = - u u_\Theta  \, ,
                                        \label{u}
\end{equation}
supplemented with the initial condition $u(0,\Theta) = t
\sin\Theta $, where, as before, $t\equiv G_T/G_D$. Employing the
method of characteristics, one may write the solution of the
latter problem in the  implicit form:
\begin{equation}
u(\zeta,\Theta) = t \sin(\Theta - \zeta u(\zeta,\Theta))\, .
                                         \label{usolution}
\end{equation}
As a result, the function $u_0(\Theta)\equiv u(1,\Theta)$, that
has $l\gamma_l$ as it's Fourier coefficients, may be found as the
solution of the  following differential  or algebraic equations
\begin{equation}
t^2 \frac{\partial}{\partial t} \left( \frac{u_0}{t} \right) + u_0
\partial_\Theta u_0 =0\, ; \hskip 1 cm
u_0(\Theta) = t \sin(\Theta - u_0(\Theta))\, .
                                         \label{u0}
\end{equation}
The last expression solves the problem of writing the action of
the barrier and the diffusive region in the form of
Eq.~(\ref{STSD}). Note, that to derive this result we have not
used a specific form of $Q_S$ and $\Lambda_N$ (other than the fact
that $Q_S^2=\Lambda_N^2=1$). Therefore the action and the
coefficients $\gamma_l$ are equally applicable to any other setup.
The only two conditions that were employed are $G_D\gg 1$ to
disregard localization effects and perform the saddle point
calculations and $E_{\rm Th}\gg E_C$ to disregard the frequency term
in the action and justify the structure of Eq.~(\ref{STSD}). The
condition $E_{\rm Th}\gg E_C$ also allows to neglect Cooper-channel
interaction in the normal conductor, while deriving proximity
action (cf.~\cite{slf01} for the discussion of general situation).

For the later use we define $\theta=\theta(t)\equiv u_0(\pi/2)$
and $\Theta_2=\Theta_2(t) \equiv u_0(\pi)$. According to
Eq.~(\ref{u0}), these two angles satisfy $\theta = t\cos\theta$
and $\Theta_2 = t\sin \Theta_2$ correspondingly. Comparing these
relations with Eqs.~(\ref{zeroUsadel1}) and (\ref{Theta2}), we
conclude that the angles $\theta$ and $\Theta_2$ introduced here
coincide with the instanton angles introduced in section \ref{s3}.

\subsection{Physical observables}
\label{s43}

In this subsection we repeat briefly derivation presented
originally in~\cite{slf01}.

{\em Normal state conductance.} Consider for a moment the dot in
the normal state with a small applied bias $V(\tau) = V
\cos\omega_m\tau$. The dot's Green function takes the form
$Q_S=\exp\{i\Phi(\tau)\}\Lambda_N(\tau-\tau')\exp\{-i\Phi(\tau')\}$,
where $\Phi =\int d\tau V(\tau)$. Substituting such $Q_S$ in the
action and taking the second variation with respect to $V(\tau)$
at $V=0$, one finds for the linear normal state conductance
\begin{equation}
G_N=G_D\sum\limits_{l=1}^\infty l^2 \gamma_l = G_D \partial_\Theta
u_0(0) \, .
                                                \label{G}
\end{equation}
For small $\Theta$ one has $u_0(\Theta) \approx \Theta
\partial_\Theta u_0(0)$; employing Eq.~(\ref{u0}), one finds
$\partial_\Theta u_0(0) = t/(1+t)$. As expected, one obtains the
resistance addition rule
\begin{equation}
G_N^{-1} =  G_D^{-1}  + G_T^{-1}  \, .
                                                \label{G1}
\end{equation}

{\em Andreev conductance.} We shall concentrate now on the
temperature region $\Delta\gg T\gg \delta$, where the SC
$Q$--matrix may be written in the form Eq.~(\ref{Qdot}) with
$\theta_n =\pi/2$, that corresponds to infinite $\Delta$
(the leading effect of the finite $\Delta$ is to renormalize
$E_C^{-1} \to E_C^{-1} +G/2\pi\Delta$ \cite{LO83,Glazman98}).
In this case the $Q_S$ matrix is  off--diagonal
in the Nambu space and time--local. Since the $\Lambda_N$--matrix
is diagonal in the Nambu space, only even powers of
$(Q_S\Lambda_N)$ may contribute to the action. As a result, one
finds for the proximity action
\begin{eqnarray}
S_{TD}[\Phi] &=& -{G_D\over 8}   \sum\limits_{l=1}^{\infty}
\gamma_{2l} \Tr\left\{ \left( Q_S \Lambda_N \right)^{2l} \right\}
                                         \label{SCaction}\\
&=& -{G_D\over 4}
\sum\limits_{l=1}^{\infty} (-1)^l \gamma_{2l}
\sum\limits_{a=1}^{2p} \underbrace{\int\limits_0^\beta\! d\tau_1
\ldots \int\limits_0^\beta\! d\tau_{2l} }_{2l} \,
e^{2i\Phi_a(\tau_1)} G(\tau_1-\tau_2) e^{-2i\Phi_a(\tau_2)}
G(\tau_2-\tau_3)\ldots e^{-2i\Phi_a(\tau_{2l})}
G(\tau_{2l}-\tau_1)\, ,             \nonumber
\end{eqnarray}
where $G(\tau-\tau') = -iT/\sin\pi T (\tau-\tau')$ is the Green
function of the normal lead (the Fourier transform of
$\Lambda_N(\epsilon_n)$ to the time domain is
$\Lambda_N(\tau,\tau')=G(\tau-\tau')\sigma_3$). If the  external
voltage $V(\tau)$ is applied to the SC dot, the phase is to be
understood as $\Phi = \int d\tau V(\tau)$. Calculating the second
variation of the above action with respect to $V(\tau)$, one finds
for the linear (Andreev) conductance
\begin{equation}
G_A= G_D\sum\limits_{l=1}^\infty  (-1)^l (2l)^2 \gamma_{2l} = G_D
\partial_\Theta u_0(\pi/2) \, .
                                                \label{GA}
\end{equation}
Employing Eq.~(\ref{u0}) for $\Theta\approx\pi/2$ one finds
\begin{equation}
G_A=\frac{G_T \sin\theta}{1+t\sin\theta} \, ,
                                                \label{GA1}
\end{equation}
where $\theta=t\cos\theta=u_0(\pi/2)$. One therefore obtains $G_A
= G_T^2/G_D + O(t) $ for $t\ll 1$ and $G_A = G_D +O(1/t)$ for
$t\gg 1$.

\subsection{Instanton treatment of the proximity  action}
\label{s42}

To calculate the proximity action on the simplest   instanton
trajectory $\Phi_1(\tau) = \pi T\tau W $ and $\Phi_a(\tau)=0$ for
$a\in [2,2p]$, one may employ e.g. the Matsubara basis (see
Appendix~\ref{app1} for calculations in the time domain).
Employing Eq.~(\ref{W1}) (modified in the obvious way for
arbitrary $W$), one finds
\begin{equation}
(Q_S \Lambda_N)^{2l} (n,m) = (-1)^{l}\delta_{n,m}\left(
\begin{array}{cc}
\mbox{sign}(\epsilon_{n-W})\mbox{sign}(\epsilon_n) & 0\\
0& \mbox{sign}(\epsilon_{n+W})\mbox{sign}(\epsilon_n)
\end{array} \right)^l\, ;
                                     \label{QSLN}
\end{equation}
in the replica $a=1$, while $(Q_S \Lambda_N)^{2l} =(-1)^l\sigma_0$
in all  other replica. As a result, for all even $l=2k$ the action
in $a=1$ replica is not different from that in the other $2p-1$
replica and therefore does not contribute to the total action in
the replica limit $p\to 0$. On the other hand, for odd $l=2k-1$,
there are $|W|$  Matsubara components, where $a=1$ and $a\neq 1$
replica come with the opposite signs. Employing
Eq.~(\ref{SCaction}), one  finds for  action on the instanton
trajectory (in the replica limit)
\begin{equation}
{\cal S}_W(t) = - |W| {G_D\over 2} \sum\limits_{k=1}^{\infty}
\gamma_{4k-2} = -|W| \frac{G_D}{8}\left[ \int\limits_0^{\pi/2}\!\!
d\Theta u_0(\Theta) - \int\limits_{\pi/2}^{\pi}\!\! d\Theta
u_0(\Theta)\right] \, ,
                                    \label{instaction}
\end{equation}
where the last equality is a direct consequence of the definition
$u_0(\Theta)=\sum_{l=1}^\infty l\gamma_l \sin (l\Theta)$.
Employing the differential equation (\ref{u0}), one finds
\begin{equation}
t^2\frac{\partial}{\partial t} \left(\frac{{\cal S}_2}{t}\right) =
\frac{G_D}{8} \left[ \int\limits_0^{\pi/2}\!\! d\Theta
\partial_\Theta u_0^2 - \int\limits_{\pi/2}^{\pi}\!\! d\Theta
\partial_\Theta u_0^2 \right] =
{G_D\over 8}\left[2\theta^2 - \Theta_2^2 \right] \, .
                                                \label{t2Sovert1}
\end{equation}
Comparing  with Eq.~(\ref{t2Sovert}), we find an exact coincidence
of the real--space result, Eq.~(\ref{CBsupr}), and the proximity
action, Eq.~(\ref{instaction}), ${\cal S}(t)={\cal S}_1(t)$.
Notice that, since the (random) diffusive region was integrated
out upon derivation of the proximity  action, the phase of the CB
oscillations is not random. In a sense the proximity action
represents a typical diffusive region (mesoscopic fluctuations are
omitted in the derivation given above). One may therefore
calculate directly typical $Z(q)$, without resorting to the
correlation function. From the two smallest winding numbers $W=1$
and $W=2$ one finds $Z(q) \approx \exp\{ -({\cal S}_{1} +{\cal
A}(T)) \} \cos\pi q + \exp\{ -{\cal S}_2\} \cos 2\pi q + \ldots $,
in agreement with the result of the previous section.

It is worth mentioning here that the general expression for the
instanton action in the case of the normal island, derived in
Ref.~\cite{Nazarov99}, may be presented in the form similar to
Eqs.~(\ref{CBsupr}) and (\ref{instaction}):
\begin{equation}
{\cal S}_W^{N} = |W|{G_D\over 4} \sum\limits_{k=1}^\infty
\gamma_{2k-1} = |W|\frac{G_D}{8}\int\limits_0^\pi \! d\Theta\,
u_0(\Theta)
= |W| {G_D\over 16} \left[ \Theta_2^2+2t(\cos\Theta_2 + 1) \right] ,
                                                \label{Ninst}
\end{equation}
and only even $W$ are allowed for the normal case. It is easy to
see, using Eq.~(\ref{Theta2}), that in the entire interval $G_D
> G_T$ the value of the action ${\cal S}_2^{N}$ is the same as in
the tunnelling limit: ${\cal S}_2^{N} = G_T/2$. As $t=G_T/G_D$
approaches unity, a kind of phase transition
occurs~\cite{Nazarov94a}, which reveals itself as a non-analytic
behavior of ${\cal S}_2^N (t)$ at $t=1$.

\subsection{Landauer approach}
\label{s44}

Employing Nazarov's techniques \cite{Nazarov94a}, one may show
that the function $u_0(\Theta)$ is directly related to the
generating function of the transmission coefficients
\begin{equation}
{1\over 2}\, G_D u_0(\Theta) = F(\Theta)\sin\Theta \equiv
\sum\limits_{\alpha = 1}^N \frac{T_\alpha \sin \Theta} {1 -
T_\alpha \sin^2(\Theta/2)}\, ,
                                             \label{Nazarov}
\end{equation}
where $T_\alpha$ are eigenvalues of the  $N\times N$  matrix
$t^\dagger t$ describing transmission between  the dot and the
lead. Factor $1/2$ on the l.h.s. takes into account that $G_D$ is
defined for the two spin components. Performing the Fourier
transform one finds
\begin{equation}
{1\over 2}\, G_D  \gamma_l =  \frac{4
(-1)^{l+1}}{l}\sum\limits_{\alpha =1}^N e^{-2\mu_\alpha l} \,\, ,
                                              \label{z}
\end{equation}
where $T_\alpha  = \cosh^{-2} \mu_\alpha$. As a result, the
proximity action takes the compact form:
\begin{equation}
S_{TD} = - \sum\limits_{\alpha=1}^N \Tr \ln \left( 1 +
e^{-2\mu_\alpha} Q_S\Lambda_N \right)   = -{1\over 2}
\sum\limits_{\alpha=1}^N \Tr \ln \left( 1 - {T_\alpha\over 4} (Q_S
-  \Lambda_N)^2 \right)\, .
                                             \label{Efetov}
\end{equation}
In the last expression we have omitted a constant which goes to
zero in the replica limit. Exactly the same interface action is
known in the supersymmetric formalism \cite{Weidenmuller,Efetov}
(notice that here we deal with  spin $1/2$ electrons).
 One may  calculate now the action on the instanton trajectory to
find the renormalized charging energy. Employing Eq.~(\ref{z}) and
recalling that ${\cal S}_{1} = -(G_D/2)\sum_k \gamma_{4k-2}$,
one finds
\begin{equation}
{\cal S}_{1}  = -{1\over 2} \sum\limits_{\alpha=1}^N  \ln \left( 1
- \frac{T_\alpha^2}{(2 - T_\alpha)^2} \right)\, .
                                              \label{Landauer}
\end{equation}
This expression is to be compared with the one for the Andreev
conductance obtained from Eqs.~(\ref{GA}) and (\ref{z})
\cite{Beenakker92,Lambert93}:
\begin{equation}
G_A = 4 \sum\limits_{\alpha=1}^N  \frac{T_\alpha^2}{(2 -
T_\alpha)^2} \, ;
                                              \label{Beenacker}
\end{equation}
(c.f. with the normal conductance $G_N=G_D\sum_l l^2\gamma_l =
2\sum_\alpha T_\alpha$, where $2$ stays for spin). Factor $4$ on
the r.h.s. of Eq.~(\ref{Beenacker}) stems from the fact that
Andreev particles carry charge $2e$ and no spin. One may thus
identify the combination ${\cal T}_\alpha = T_\alpha^2/(2 -
T_\alpha)^2$ as the Andreev conductance in channel $\alpha$. The
renormalisation of the charging energy is therefore given by
\begin{equation}
e^{-{\cal S}_{1} } = \prod\limits_{\alpha=1}^N (1-{\cal
T}_\alpha)^{1/2} \, .
                                            \label{us}
\end{equation}
This may be compared with the corresponding factor for the normal
{\em spinless} particles \cite{Nazarov99}:  $\prod_\alpha(1 -
T_\alpha)^{1/2}$. Indeed, in our notations  the normal instanton
action is ${\cal S}_2^N = (G_D/2) \sum_l \gamma_{2l-1} =
-\sum_\alpha\ln(1-T_\alpha)$; for spinless particles one has
$\exp\{-{\cal S}_2^N/2\}$ for the CB suppression. There is thus a
perfect analogy between Andreev and normal spinless charge
fluctuations mechanisms, provided that the normal transmission
coefficients, $T_\alpha$, are substituted by the Andreev ones,
${\cal T}_\alpha$.

{}From Eq.~(\ref{us}) one may extract some general results. In the
tunnelling limit: ${\cal T}_\alpha \ll 1$ in all channels, one
finds: ${\cal S}_{1} = G_A/8$. In the diffusive metal interface
the transmission eigenvalues are distributed according to the
Dorokhov distribution \cite{Dorokhov,Nazarov94a}:
$P(T)=G_D/(4T\sqrt{1-T})$. The corresponding distribution for the
Andreev transmissions is given by $P({\cal T}) = G_D/(8{\cal T}
\sqrt{1 - {\cal T}})$, as a result $G_A=G_D$. For the typical
action of the diffusive interface one finds therefore $\langle
{\cal S}_{1} \rangle = \pi^2 G_A/32$.

\subsection{Fluctuation determinant and summation over many-instanton
configurations}

We now expand the action, Eq.~(\ref{SCaction}), to the second
order in deviations from the instanton trajectory $\Phi_a(\tau) =
\pi T \tau W_a + \delta\Phi^a(\tau)$, where $\delta\Phi^a(0) =
\delta\Phi^a(\beta)$. After diagonalization of the resulting
quadratic form one finds the following spectrum of the small
fluctuations \cite{Zaikin,Grabert,Andreev}:
\begin{equation}
\lambda_n^{(W_a)} = \frac{\pi^2 T}{E_C}\, n^2 + \frac{G_A}{4}
\left( |n - W_a| + |n+ W_a| - 2|W_a| \right) \, ,
                                               \label{spectrum}
\end{equation}
where $n\in [-\infty,\infty]$. For $T\ll E_C$ there are $2|W_a|
+1$ (almost) zero modes labelled by $n\leq |W_a|$. One of them
$n=0$ is the trivial shift of $\Phi_a(\tau)$ by a constant,
whereas the remaining $2|W_a|$ zero modes are associated with the
deformation of the instantons. The general  solution of the saddle
point equations (for $T\ll E_C$) known as the Korshunov
instanton~\cite{Korshunov}  may be written as
\begin{equation}
e^{2i\Phi_a(\tau)} = \prod\limits_{k=1}^{|W_a|} \left[
\frac{e^{2\pi iT\tau} -z_k}{1-  e^{2\pi iT\tau} z_k^*}
\right]^{\mathop{\rm sign}(W_a) } \, ,
                                     \label{korshunov}
\end{equation}
where $z_k$ is a set of $|W_a|$ complex numbers (instanton
coordinates) which parametrize the $2|W_a|$ dimensional zero--mode
manifold. The instanton coordinates, $z_k$, are complex numbers
from  inside of the unit circle. One may thus characterize any
deviation $\delta\Phi_a(\tau)$ by $2|W_a|$ zero--mode coordinates,
$z_k$, and remaining transversal modes having non--zero masses,
Eq.~(\ref{spectrum}). The Jacobian of such transformation may be
found in the standard way \cite{Zinn-Justin} and is given by
\begin{equation}
J = {1\over |W_a|!}\, \mbox{det}\left|\left| {1\over 1 - z_l
z_k^*} \right|\right| \, ,
                                              \label{jacobian}
\end{equation}
where $l,k \in [1,|W_a|] $. The fact that the Jacobian vanishes if
two of the  coordinates coincide signals the repulsive interaction
between the instantons.

The Gaussian integration over the massive degrees of freedom
normalized by those in the zero winding number sector results in
\begin{equation}
\frac{\prod\limits_{n=  1}^{\infty} (\lambda_n^{(0)}/\pi) }
{\prod\limits_{n=  |W_a|+1}^{\infty}\!\!\!\!\!\!\!
(\lambda_n^{(W_a)}/\pi) } = \left( {G_A\over 2\pi} \right)^{|W_a|}
\exp\left\{|W_a| \ln \left({G_A E_C\over 2\pi^2T} \right)\right\}
\, ,
                                                 \label{fluctdet}
\end{equation}
up to the terms of the order $T/(G_A E_C)\ll 1$. Combining
together all the factors, one finds for the one instanton,
$W_a=1$, trajectory
\begin{equation}
\frac{Z_1}{Z_0} = e^{i\pi q} \oint\limits_{|z|\leq 1} \!\! \frac
{d^2 z}{1-|z|^2}\, \left({G_A\over 2\pi}\right) \, e^{-\{{\cal
S}_1 - \ln(G_A E_C/2\pi^2 T)\}} = e^{i\pi q}\,\, {\tilde E_C \over
2 T}\, \ln {E_C\over T}\, ,
                                           \label{singleinst}
\end{equation}
The cutoff of the logarithmically divergent $z$--integral
originates from the following consideration. In presence of the
charging energy term the Korshunov instantons,
Eq.~(\ref{korshunov}), are not true zero--modes of the action. The
$z$--dependence  of the action brings the  factor
$\exp\{-(T/2E_C)|z|^2/(1-|z|^2)\}$ to the integral. As a result
the effective integration range shrinks to $|z|^2 < 1-T/E_C<1$.
The renormalized charging energy $\tilde{E}_C$ is defined as
\begin{equation}
\tilde E_C = {E_C G_A^2\over 2\pi^2}\,  e^{-{\cal S}_1}\, .
                                         \label{tilde}
\end{equation}
 It is also instructive to look at
the $W_a=2$ contribution to the  partition function
\begin{eqnarray}
\frac{Z_2}{Z_0} &=& e^{2i\pi q} {1\over 2!}   \oint\!\!\oint \!\!
d^2 z_1 d^2 z_2 \left[ \frac{1}{(1-|z|_1^2)(1 - |z|^2_2)} -
\frac{1}{(1 - z_1 z_2^*)(1 - z_1^*z_2)} \right] \, \left(
{G_A\over 2\pi} \right)^2\,
e^{-\{{\cal S}_2 - 2\ln(G_A E_C/2\pi^2T)\}} \nonumber     \\
 &=&
  {1\over 2!}\; e^{2i\pi q}\, \left(  {\tilde E_C\over 2 T}\right)^2
  \left[\ln^2 {E_C\over T} - {\pi^2\over 6}\, \right] \, .
                                           \label{doubleinst}
\end{eqnarray}
The leading power of the large logarithm comes from the diagonal
term of the Jacobian, Eq.~(\ref{jacobian}). This term corresponds
to the non--interacting instantons approximation, which is
justified therefore by the large parameter $\ln (E_C/T)\gg 1$.
Notice, that the (repulsive) interaction correction (the
off--diagonal part of the Jacobian, Eq.~(\ref{jacobian})) comes
with $\ln^0$ and not $\ln^1$, as may be naively expected.

Due to the large logarithmic factor, originating from the
integration over sizes of each individual instanton (i.e.
neglecting their interactions), one may treat  positive  and
negative instantons (anti--instantons) on the equal footing.
Strictly speaking, a combination of instantons and
anti--instantons is not a saddle point solution. However, relative
weakness of instanton interactions (no log-factor in the second
term in Eq.~(\ref{doubleinst})) makes it possible to consider an
ideal gas of instantons and anti--instantons.
The contribution to the partition function with a given winding number
$W$ is given by all configurations having $m+|W|$ instantons and
$m$ anti--instantons in an arbitrary order. Taking only the terms
with the  leading power of the logarithms (diagonal term of the
Jacobian) one finds \cite{Grabert}
\begin{equation}
Z(q) = Z_0\sum\limits_{W=-\infty}^{\infty} e^{\pi i W q}\,
\sum\limits_{m=0}^{\infty} {1\over m! (m+|W|)! } \left( {\tilde
E_C\over 2 T}\;  \ln {E_C\over T} \right)^{2m+|W|}\, ,
                                          \label{gas}
\end{equation}
This ugly looking series may be summed up into an unexpectedly
simple expression:
\begin{equation}
Z(q) = Z_0\exp\left\{  {\tilde E_C\over T}\;   \ln\left( {E_C\over
T}\right)  \cos (\pi q) \right\}\, .
                                          \label{freee}
\end{equation}
The simplest way to check it is to expand back Eq.~(\ref{freee})
in a double series in powers of $\tilde E_C $ and $e^{i \pi q}$.
The role of the anti--instantons, therefore, is to convert $\cos
(W \pi q)$, which may be expected from the instantons only, into
$\cos^W(\pi q)$.

Under the condition $\tilde{E}_C \leq T \ll E_C$ the interaction
between instantons can be treated perturbatively; for the
lowest-order correction we use the last term in
Eq.~(\ref{doubleinst}) to find for the gate voltage - dependent
free energy
\begin{equation}
F(q) = - \tilde E_C \left[ \ln \left({E_C\over T}\right)  \cos(\pi
q) - {\pi^2\over 24} {\tilde E_C\over T} \cos (2\pi q)+ \ldots
\right] \, .
                                                    \label{freeenergy}
\end{equation}
We retain the second term in (\ref{freeenergy}) since it will be
important at temperatures comparable with the parity-effect
temperature $T^*$, see Sec.~\ref{s7}. It also demonstrates that
the instanton--instanton interaction corrections are small only if
$\tilde E_C< T$.

\section{RG treatment of the quantum fluctuations}
\label{s5}

We next approach the problem of the CB from the different
perspective. Instead of calculating the  action  on the instanton
field configurations, we shall perform the  RG analysis of the
coefficients $\gamma_l$ upon integrating out  fast fluctuations of
$\Phi(\tau)$. To this end we write $\Phi(\tau) = \Phi^{s}(\tau) +
\Phi^f(\tau)$, where superscripts $s$ and $f$ stay for slow
($\omega_m < \Omega$) and fast ($\omega_m > \Omega$) component of
the field correspondingly. The running cutoff $\Omega$ runs from
$\Omega \sim G_A E_C$ down to the lowest Matsubara frequency
$\Omega\sim T$. We next substitute the $\Phi$--field into the SC
proximity action, Eq.~(\ref{SCaction}), expand to the second order
in $\Phi^f$ and integrate out the fast field fluctuations. The
bare propagator of the $\Phi^f$--fields may be read out from the
action Eq.~(\ref{SCaction}):
\begin{equation}
\langle \Phi^f_a(\omega_m) \Phi^f_b(-\omega_m)\rangle =
\frac{\delta_{ab}}{|\omega_m| G_A(\zeta_\Omega) }\, ,
                                           \label{propagator}
\end{equation}
where $G_A(\zeta_\Omega) = G_D\sum_{l=1}^{\infty} (-1)^l (2l)^2
\gamma_{2l} (\zeta_\Omega)$ is the running value of the Andreev
conductance, and $\zeta_\Omega \equiv \ln (G_A E_C)/\Omega$. (Note
that this definition of $\zeta$ differs from that used to derive
the proximity action in Section~\ref{s4}.) The further calculation
is essentially similar to that in the Keldysh
formalism~\cite{fls01}. One obtains the following set of the
evolution equations for $\gamma_l(\zeta_\Omega)$:
\begin{equation}
\frac{d\gamma_{2l}}{d\zeta} = -\frac{8}{ G_A(\zeta)} \left(
l\gamma_{2l} + 2\sum\limits_{k=1}^\infty (-1)^k (l+k)
\gamma_{2l+2k} \right) \, ,
                                         \label{RG}
\end{equation}
where we have omitted the subscript $\Omega$. Since only the even
coefficients are involved it is convenient to define the function
$\tilde{u}(\zeta,\Theta) =  [u(\zeta,\Theta) -
u(\zeta,\pi-\Theta)]/2$ which contains only even harmonics. In
terms of this function the RG equations (\ref{RG}) take the form
\begin{equation}
  \tilde{u}_\zeta = - \frac{4}{ G_A(\zeta)}\left[\tilde{u}(\Theta)\tan
  \Theta\right]_\Theta\, ,
                                                         \label{simple}
\end{equation}
where the Andreev conductance is $G_A(\zeta) =G_D
\tilde{u}_\Theta(\zeta,\pi/2)$. The initial condition for
Eq.~(\ref{simple}) is  $\tilde{u}(0,\Theta) =
\tilde{u}_0(\Theta)$, with the $u_0(\Theta)$ function evaluated
above. Let us mention for  completeness, that in the normal case
the corresponding RG equation takes the form $u_\zeta =
(2/G_N(\zeta)) [u(\Theta)\cot( \Theta/2)]_\Theta$, where the
running normal conductance according to Eq.~(\ref{G}) is given by
$G_N(\zeta) = G_D\sum_{l=1}^{\infty} l^2 \gamma_{l} (\zeta)=G_D
u_\Theta(\zeta,0)$.

The key parameter that determines the strength of phase
fluctuations is the Andreev conductance $G_A(\zeta)$. We consider
the large bare value, $G_A(0) \gg 1$.  Upon integration over
fluctuations of $\Phi(\tau)$, the effective value of $G_A(\zeta)$
decreases (the physical reason  of this phenomena is a partial
loss of the phase coherence between multiple Andreev reflections),
and becomes comparable  to unity at some time scale
$\Omega_c^{-1}=(G_AE_C)^{-1}e^{\zeta_c}$.  At longer times
coupling between the dot and the lead  is  weak, thus the  phase
fluctuations grow and autocorrelation function $C(\tau) =\langle
e^{i(\Phi(\tau)-\Phi(0))}\rangle$ decays fast. It is natural to
associate $\Omega_c$ with an effective Coulomb energy of the
island connected to the wire, $\tilde{E_C}$. We will show now that
the estimate for $\Omega_c$ that follows from the RG equation
coincides (within exponential accuracy) with the results of the
instanton analysis.

Equation (\ref{simple}) may be solved with the method of
characteristics. We define the  function $v(\Theta) =
\tilde{u}(\Theta)\tan{\Theta}$ and then introduce new ``space'' and
``time'' variables: $\xi = -\ln \sin{\Theta}$ and $\tau$ which is
defined via the relation
\begin{equation}
  \frac{4}{ G_A(\zeta)}d\zeta = d\tau\, .
                                                     \label{tau}
\end{equation}
In terms of these new variables the RG equation (\ref{simple})
takes the simple form: $v_\tau = v_\xi$, yielding $v(\xi,\tau) =
v_0(\xi+\tau)$. As a result,
\begin{equation}
  \tilde{u}(\zeta(\tau),\Theta)
  = \frac{e^{-\tau}\cos \Theta  }
    {\sqrt{1-e^{-2\tau}\sin^2{\Theta}}}\; \tilde{u}_0(\arcsin(e^{-\tau}\sin{\Theta}))\, ,
                                                                 \label{solu}
\end{equation}
The dependence $\zeta(\tau)$ can be now found from Eq.~(\ref{tau})
with the  Andreev conductance given by $G_A(\zeta) =
G_D\tilde{u}_\Theta(\zeta,\pi/2)$.

Our goal is  to find a scale $\zeta_c$, where the Andreev
conductance $G_A$ becomes small. To find $\zeta_c$ we notice  that
in terms of the RG time $\tau$ the smallness of $G_A$ means the
limit $\tau \to \infty$. Then, using definition of $\tau$ given by
Eq.~(\ref{tau}), we find
\begin{equation}
  \zeta_c
  = - \frac{G_D}{4} \int\limits_0^\infty\!\! d\tau\, \frac{\tilde{u}_0(\arcsin{e^{-\tau}}) }
    {\sqrt{e^{2\tau} - 1}}
  = - \frac{G_D}{4} \int\limits_0^{\pi/2}\!\!
  d\Theta\, \tilde{u}_0(\Theta)  \, .
                                                          \label{otvet}
\end{equation}
Recalling the definition of $\tilde u_0(\Theta)$ and comparing
this result with Eq.~(\ref{instaction}) we conclude that $\zeta_c
= {\cal S}_{1} $. As a result, the amplitude of the lowest
oscillatory component ($\sim \cos\pi q$) of the free energy, from
the RG point of view, is given by  the precisely the same
exponential factor as in the framework of the instanton
calculation.

Yet another instructive way to approach the problem is to follow
the renormalization of the instanton action. According to
Eq.~(\ref{instaction}), ${\cal S}_W(\zeta) = -|W| (G_D/2)
\sum_k \gamma_{4k-2}(\zeta) = -|W| (G_D/4)
\int_0^{\pi/2}\!\! d\Theta\, \tilde{u}_0(\Theta)$. Employing
Eq.~(\ref{simple}), one finds
\begin{equation}
\frac{d {\cal S}_W }{d \zeta} = \frac{|W| G_D}{G_A(\zeta)}
\left( \tilde{u}(\Theta) \tan \Theta \right)\Bigr|_{\Theta\to
{\pi\over 2} } =  - |W| \, ,
                                                  \label{Srenorm}
\end{equation}
where in the last equality we have used that $G_A(\zeta) =
G_D\tilde{u}_\Theta(\zeta,\pi/2)$. (In the normal case one finds
$dS_W^N/d\zeta = -|W|/2$; only even $W$ are allowed.)  From the
dimensional analysis one expects $F(q) \sim \Omega \exp\{-{\cal
S}_{1} \} \cos\pi q +\ldots $, where $\Omega\sim E_C G_A$ is the
cutoff energy. We may now renormalize down the cutoff energy
$\Omega \to \Omega(\zeta) \equiv \Omega e^{-\zeta}$,
simultaneously following renormalization of the action, ${\cal
S}_{1} \to {\cal S}_{1}(\zeta)$. According to Eq.~(\ref{Srenorm}),
${\cal S}_{1}(\zeta) = {\cal S}_{1} - \zeta$. As a result,
\begin{equation}
\Omega(\zeta)\, e^{-{\cal S}_{1}(\zeta) } = \mbox{const} \, .
                                          \label{const}
\end{equation}
This means that the lowest oscillatory component of the free
energy is expected to be temperature independent, in agreement
with Eq.~(\ref{freeenergy}) (up to log--factor, originating from
the zero modes).
 The instanton calculation is done at $\zeta =0$,
alternatively the RG calculation presented above looks for
$\zeta_c$, such that ${\cal S}_{W}(\zeta_c) \approx 0$ (the scale
where the instantons do not cost anything). In view of
Eq.~(\ref{const}) it is not surprising that they yield the same
result. Moreover, one can renormalize down to any intermediate
scale $\zeta$ (such that $G_A(\zeta) > 1$) and then calculate the
instanton action with the current parameters $\{\gamma_l(\zeta)\}$
-- the result is still guaranteed to be the same. Notice also that
the renormalized action $S_W(\zeta) = S_W - |W|\ln (G_A E_C/T)$
coincides with the result that comes from the instanton action
together with the massive Gaussian fluctuations, c.f.
Eqs.~(\ref{fluctdet})--(\ref{doubleinst}). The effect of the zero
modes ($\ln E_C/T$ in the prefactor) is missed  in the one--loop
RG calculation.

Finally we comment upon the comparison of the approach used in this Section
 with  Ref.~\cite{fls01} where the same kind of RG approach was
employed in the case of 2-dimensional normal conductor with very
 low Thouless energy, $E_{\rm Th} \ll E_C$. In the last case
the RG procedures involves simultaneous integration over phase
fluctuations $\Phi(\tau)$ {\it and} diffuson/Cooperon modes in the
normal conductor. As a result, the full functional RG equation
contains three terms in the r.h.s.: one coming from Eq.~(\ref{u}),
the second is similar to Eq.~(\ref{simple}), and the third
contribution accounts for the Cooper-channel repulsion in the
normal conductor (cf.~\cite{slf01,slf01short}).
 The resultant RG
equation cannot be reduced to the set of even harmonics
$\tilde{u}$ only; we are not aware of any method to solve it apart
from ``brut-force'' numerical integration.  Considerable
simplification of the problem treated here stems from the fact
that, due to the condition $E_{\rm Th} \gg E_C$ all diffuson/Cooperon
modes may be  integrated out {\it before} (means at larger
energies than) the phase fluctuations $\Phi(\tau)$ become
relevant. This procedure leads to the action functional $S_{TD}$
that depends upon $\Phi(\tau)$ trajectory only, as defined in
(\ref{SCaction}). Thus the results presented in this Section can
be considered as ``minimal generalization'' of those obtained in
Ref.~\cite{fl98}.

\section{Discussion of the results}
\label{s7}

The main result of our study,  given by Eq.~(\ref{freeenergy}),
applies in the temperature range
\begin{equation}
 \tilde{E}_C \leq T \ll \min\{E_C, T^*\}\, ,
                                                           \label{window}
\end{equation}
where $T^*$ is the parity-effect temperature given by
Eq.~(\ref{Tstar}). The condition  $T\ll T^*$ ensures that both
even and odd winding numbers contribute to the partition function,
since  ${\cal A}(T) \ll 1$, c.f. Eq.~(\ref{diff}). The two lowest
oscillatory components of the free energy are given then by
Eq.~(\ref{freeenergy}). At higher temperature $T\sim T^*$ an
addition of odd number of electrons to the dot becomes possible,
changing the relative amplitude of the harmonics. On the level of
the partition function the components with odd winding numbers
acquire the temperature dependent factor $\exp\{-{\cal A}(T)\}$.
As a result, one finds for the gate voltage dependent free energy:
\begin{equation}
F(q) = - \tilde E_C \left[ e^{-{\cal A}(T)} \ln\left( {E_C\over
T}\right) \cos(\pi q)  +  \left(   ( 1 - e^{-2{\cal A}(T)})\ln^2
\left({E_C\over T}\right)  - {\pi^2\over 6}\right)  {\tilde
E_C\over4 T} \cos (2\pi q)+ \ldots \right] \, .
                                                    \label{freeenergy1}
\end{equation}
Finally at $T > T^*$ the parity effect disappears and only the
normal (unit) oscillation period remains in the free energy
\begin{equation}
F(q) = - \tilde E_C    {\tilde E_C\left( \ln^2\left( {E_C\over
T}\right) - {\pi^2\over 6}\right) \over 4 T} \cos (2\pi q)+ \ldots
                                                    \label{freeenergy2}
\end{equation}
Notice that if $\ln E_C/T^* > \pi/\sqrt{6}$ there is the sign
change of the $\cos(2\pi q)$ component at $T\approx T^*$.
Moreover, unlike the low temperature case, where the amplitude of
the dominant harmonics was weakly (logarithmically) temperature
dependent, Eq.~(\ref{freeenergy}), at larger temperature there is
the stronger dependence ($\sim T^{-1} \ln^2(E_C/ T)$) of the
amplitude. The characteristic crossover temperature is determined
from the relation ${\cal A}(T^{\dagger}) \approx \ln
(T^{\dagger}/\tilde E_C)$ and is given by
\begin{equation}
T^{\dagger} = \frac{\Delta}{\ln {\Delta \over G_A \delta}}\, ,
                                             \label{Tstar1}
\end{equation}
where we have used the fact that ${\cal S}_{1} \sim G_A$.

We  emphasize an interesting feature of the weak Coulomb blockade
compared to the usual one: although the role of effective Coulomb
energy is taken by the effective energy $\tilde{E}_C \ll E_C$, the
oscillation amplitude depends relatively weakly upon temperature
within the range given by Eq.~(\ref{window}). This is to be
contrasted with the usual case of an island connected by highly
resistive contacts:
 at temperatures $ T > E_C$ oscillation amplitude vanishes exponentially
fast with $T/4E_C$.

\begin{figure}
\epsfxsize=8cm
\centerline{\epsfbox{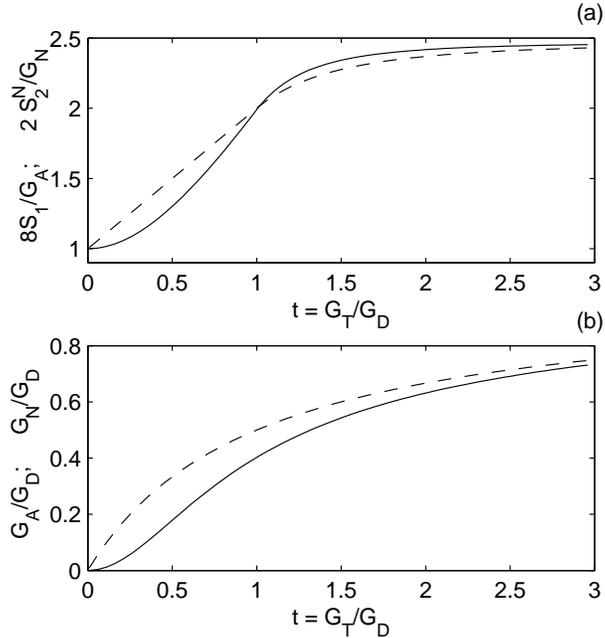}}
\vspace{2mm}
\caption{(a) Normalized actions $8{\cal S}_1/G_A$ (full line) and
$2{\cal S}_2^N/G_N$ (dashed line) as functions of $t=G_T/G_D$.
Both graphs interpolate between 1 in the tunnelling limit and $\pi^2/4$
in the diffusive limit.
(b) Andreev (full line) and normal (dashed line) conductances normalized
by $G_D$ as functions of $t$.}
\label{fig2}
\end{figure}

\begin{figure}
\epsfxsize=8cm
\centerline{\epsfbox{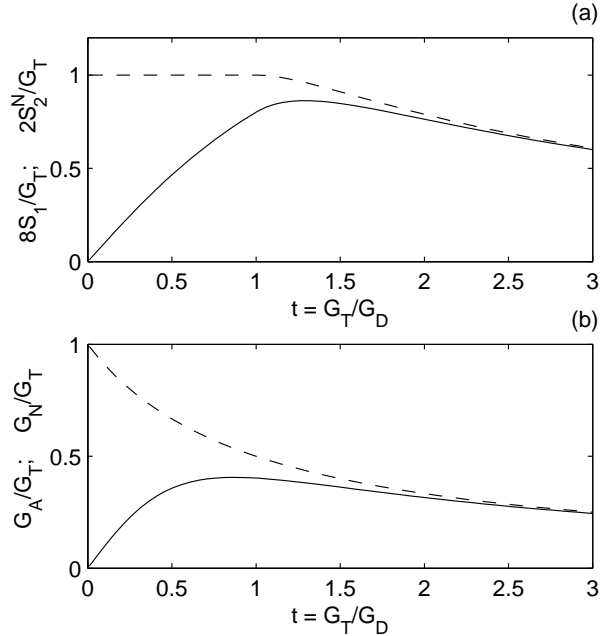}}
\vspace{2mm}
\caption{The same as Fig.~\protect\ref{fig2} but normalized by $G_T$.
Notice that the normal action exhibits a continuous ``phase transition''
at $G_D=G_T$. The Andreev action has a maximum at $G_D\approx 0.8G_T$
which is a direct consequence of nonmonotonous dependence of the Andreev
conductance on the resistance of the diffusive metal.}
\label{fig3}
\end{figure}

Next we shall discuss the quantitative value of the renormalized
charging energy $\tilde E_C$, which determines both the magnitude
of the CB and the region of applicability of our results. To this
end we need to evaluate ${\cal S}_{1}(t)$ (where $t=G_T/G_D$),
which is given by one of the two equivalent expressions,
Eqs.~(\ref{CBsupr}) or (\ref{instaction}). In the two limiting
cases $t\ll 1$ (the tunneling barrier limit) and $t\gg 1 $ (the
diffusive metal limit) the answer may be found analytically. One
obtains
 \begin{equation}
 {\cal S}_{1} = -\ln \frac{2\pi^2\tilde{E}_C}{G_A^2 E_C} = \frac{G_A}{8}
  \cases{
 1\, ; & in the tunneling limit, \cr
 \pi^2/4\, ; & in the diffusive limit,
}
                                               \label{final1}
\end{equation}
where the Andreev conductance is given by Eq.~(\ref{GA1}). Recall
that for a normal dot the corresponding exponent is given by
$G_N/2$ in the tunnelling limit and $G_N\pi^2/8$ in the diffusive
limit. We observe that upon the same dissipative conductance the
SC dot exhibits factor of four (in the two limiting cases) smaller
action. Therefore one may observe a sizeable CB in the
superconductive state, while in the analogous normal dot the CB is
practically suppressed. The crossover behavior for the SC action
${\cal S}_1(t)$, the normal action ${\cal S}_2^N(t)$ and the
Andreev conductance $G_A(t)$ as function of $t=G_T/G_D$ is
depicted in Fig.~\ref{fig2}.
Dependence of the same quantities on the resistance $G_D^{-1}$
of the normal region at fixed $G_T$ is demonstrated in Fig.~\ref{fig3}.

In the case $\tilde{E}_C \ll T^{\dagger}$ the crossover between $T
< T^{\dagger}$ and $T > T^{\dagger}$  regimes leads to the sharp
drop of the oscillation amplitude by the factor
$\frac{\tilde{E}_C}{4T^{\dagger}}\ln\frac{E_C}{T^{\dagger}} \ll 1$
The same drop in the residual Coulomb blockade may occur as
function of magnetic field, due to the suppression of the $\Delta$
and thus of $T^{\dagger}$ by the magnetic field.  This may result
in a strong negative magnetoresistance at low temperatures  of a
granular media made out
 of small superconductive grains
(cf. e.g.~\cite{mesointro} for the discussion of relevant experiments).
This effect can occur if the low-temperature oscillation amplitude
$\tilde{E}_C$ is sufficiently large to destroy Josephson coupling
 between grains, so transport of Cooper pairs between grains is blocked
by $\tilde{E}_C$. However,  quantitative theory of such an
effect is still to be developed.

Another open question concerns the behavior of $F(q)$ in the zero
temperature limit, $ T < \tilde{E}_C$. As temperature decreases
below $\tilde{E}_C$, all approximate methods, we used, run out of
their applicability range: the renormalized conductance $G_A$
drops below unity, fluctuations become strong and there is no {\em
apriori} reason to treat them within Gaussian approximation.
Whereas the overall scale of the oscillation amplitude  is
probably given by its first harmonic, Eq.~(\ref{freeenergy}) at
$T\sim \tilde E_C$, and is of the order $\tilde E_C \ln {E_C\over
\tilde E_C}$, the precise shape of the oscillations is still to be
determined.
It is possible that in the $T\to 0$ limit the function $F(q)$
becomes nonanalytic at the degeneracy point $q=1/2$.
An extreme case of such a nonanalytic behavior -- finite steps in
$dF/dq$ at half-integer $q$ -- was found in Ref.~\cite{Glazman98}
where the Andreev conductance was completely neglected.

\acknowledgments

We have greatly benefitted from the numerous discussions with
A.~V.~Andreev, I.~S.~Beloborodov, L.~I.~Glazman, and K.~A.~Matveev.
MVF and MAS were supported by the
SCOPES program of Switzerland, Dutch Organization for Fundamental
Research (NWO), Russian Foundation for Basic Research under grant
01-02-17759, the program ``Quantum Macrophysics'' of the Russian
Academy of Sciences, the Russian Ministry of Science, and the
Swiss National Foundation. AK was partially supported  by the BSF
grant N  9800338. AIL was partially supported by the NSF grant N
0120702. MAS was partially supported by the Russian science support
foundation.

\appendix

\section{Instantons in the Imaginary-Time Proximity Action}
\label{app1}

\subsection{Transition to imaginary time}

The method of the multicharge proximity action was initially
developed~\cite{slf01,slf01short} in the Keldysh real-time
representation. Here we show how the Matsubara proximity action,
Eq.~(\ref{STSD}), may be obtained from its Keldysh analog by
analytic continuation to imaginary time. Since this Appendix
serves illustrative purposes we will consider the simplest case
$T=0$. As discussed in the bulk of the paper, the form of $F(q)$
is unknown at $T=0$ since at $T<\tilde E_C$ the dilute instanton
gas approximation fails and one has to consider an interacting
instanton liquid. Nevertheless, the overall scale of oscillations
in $F(q)$ may be inferred from the single instanton action which
is temperature independent and may be calculated at $T=0$.

To proceed we need to establish a correspondence between our notations and
those adopted in Refs.~\cite{fls00,slf01,slf01short}. The latter notations
will be designated by a prime. Firstly, our conductance quantum
$G_Q=e^2/2\pi\hbar$ is different from $G'_Q=e^2/\hbar$ used in those papers.
Secondly, the running parameter $\zeta$ and charges $\{\gamma_n\}$ are
related by
\be
  \frac{\zeta}{\zeta'} = \frac{\gamma_n'}{\gamma_n} = \frac{G_D}{8\pi^2g'} .
\ee
Thus, the Keldysh multicharge proximity action may be written
in our notations as~\cite{actions}
\be
  S_{\rm prox}[\Phi(t)]
  = -i \frac{G_D}{8} \sum_{n=1}^\infty \gamma_{n}(\zeta)
    \Tr (\check Q_S \check\Lambda_N)^{n}\, ,
\label{sg}
\ee
where the trace is taken over time indexes, Nambu
and Keldysh spaces. Here $\check Q_S$ is a matrix in the
superconducting dot acting in the Nambu space as
\be
  \check Q_S(t) =
  \left( \begin{array}{cc}
    0 & e^{2i\tensor\Phi(t)} \\
    e^{-2i\tensor\Phi(t)} & 0
  \end{array} \right) ,
\label{QS}
\ee
where $\tensor\Phi = \mathop{\rm diag}(\Phi_>,\Phi_<)$ is a matrix
in the Keldysh space, with $\Phi_>(t)$ and $\Phi_<(t)$ being the fields
residing on the forward and backward branches of the Keldysh contour.
In the normal lead, $\check\Lambda_N(E) = \check\Lambda_0(E)
\otimes \sigma_3$, where $\sigma_3$ is the Pauli matrix in the Nambu
space, and the matrix $\check\Lambda_N(E)$ acts in the Keldysh space:
\be
  \check \Lambda_0(E) =
  \left( \begin{array}{cc}
    1-2f & -2f \\
    2(f-1) & 2f-1
  \end{array} \right) ,
\ee
$f(E)$ being the distribution function.

Tracing over the Nambu space reduces the action~(\ref{sg}) to the form:
\be
  S_{\rm prox}[\Phi(t)]
  = -i \frac{G_D}{4} \sum_{n=1}^\infty (-1)^n \gamma_{2n}
  \int\limits_{-\infty}^\infty dt_1 \dots dt_{2n} \mathop{{\rm tr}_K}
    e^{2i\tensor\Phi(t_1)} \check\Lambda_0(t_1-t_2)
    e^{-2i\tensor\Phi(t_2)} \check\Lambda_0(t_2-t_3)
    \dots
    e^{-2i\tensor\Phi(t_{2n})} \check\Lambda_0(t_{2n}-t_1)
  ,
\label{prox-s}
\ee
where the trace is taken only over the Keldysh space.

\begin{figure}
\centerline{\epsfbox{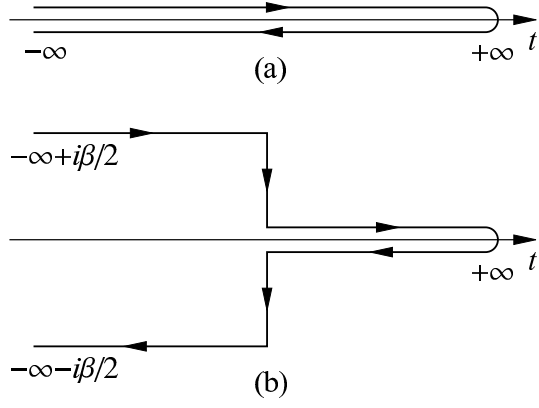}}
\vspace{2mm}
\caption{The initial Keldysh contour (a) and its deformation to imaginary times (b).}
\label{fig4}
\end{figure}

Transition to the imaginary time is achieved by the deformation of
the Keldysh contour $\cal C$ which initially run over the
time axis from $-\infty$ to $\infty$ and then in the backward
direction. The desired deformation introduces a vertical segment
of length $\beta$ at some time $t=t_0$ so that the contour
originates at $-\infty+i\beta/2$, runs through the points
$t_0+i\beta/2$, $t_0$ to $\infty$ and then back through $t_0$,
$t_0-i\beta/2$ to $-\infty-i\beta/2$, see Fig.~\ref{fig4}.
For the purpose of the
evaluation of the instanton action the choice of the point $t_0$
is somewhat arbitrary. Note, however, that were we to consider,
e.g., the correlation function of the form $\langle\tilde N(q)
\tilde N(q')\rangle$, it would be necessary to allow for adiabatic
evolution of the gate voltage $q=q(t)$, $q'=q(t')$ and to
introduce two vertical segments at times $t$ and $t'$.

The analysis is simplified in the zero-temperature case.
Choosing the position of the vertical segment at $t_0=0$, we deform
the forward segment $(-\infty,0)$ of the Keldysh contour
to the upper imaginary half-axis
and the backward segment $(0,-\infty)$
to the lower imaginary half-axis.
The remaining horizontal appendix of the Keldysh contour running
from 0 to $\infty$ and then back to $t=0$ can be neglected as it
does not contribute to thermodynamic quantities.

In order to continue the function $\check\Lambda_0(t)$
to the complex plane in time
one should employ its analytic properties~\cite{analyt}.
In deforming the contour to the imaginary axis
the element $\check\Lambda_0^{12}(\tau)$ enters only with $\tau>0$,
and the element $\check\Lambda_0^{21}(\tau)$ enters only with $\tau<0$.
Under this condition the function $\check\Lambda_0(\tau)$ can be
substituted by~\cite{LO83}
\be
\label{g}
  \check\Lambda_0(\tau) \to F(\tau)
  \left( \begin{array}{cc}
    1 & 1 \\
    -1 & -1
  \end{array} \right) ,
\qquad
  F(\tau) = - \frac{1}{\pi\tau}.
\ee
Eq.~(\ref{g}) solves the problem of analytic continuation of the Green
function $\check\Lambda_0(t)$.

The trivial matrix structure of Eq.~(\ref{g}) ensures that each term
in Eq.~(\ref{prox-s}) can be written as a multiple integral
of the single function
\be
  \Phi(\tau) =
  \cases{
    \Phi_>(\tau), & for $\tau>0$, \cr
    \Phi_<(\tau), & for $\tau<0$,
  }
\ee
defined on the whole imaginary axis.
The infinitesimal element $dt$ is transformed to
\be
  dt \to
  \cases{
    -i\,d\tau, & for $\tau>0$, \cr
     i\,d\tau, & for $\tau<0$.
  }
\ee
Finally, we obtain~\cite{actions}
\be
  S_{\rm prox}[\Phi(\tau)] =
  -\frac{G_D}{4} \sum_{n=1}^\infty \gamma_{2n}
  \int_{-\infty}^\infty d\tau_1 \dots d\tau_{2n} \:
    e^{2i\Phi(\tau_1)-2i\Phi(\tau_2)+\dots-2i\Phi(\tau_{2n})} \:
    F(\tau_1-\tau_2) F(\tau_2-\tau_3) \dots F(\tau_{2n}-\tau_1) ,
\label{S-imag}
\ee
that coincides with Eq.~(\ref{SCaction}) in the zero-temperature limit.

\subsection{Instantons in time domain}

Here we calculate the action on the trajectories for which
$e^{2i\Phi(\tau)}$ is an analytic function in the upper
half-plane. For such a solution, the instanton's winding number
$\pi W = \Delta\Phi = \int d\tau (\partial\Phi/\partial\tau)$ is
positive. Integration over $t_1$, $t_3$, \dots, $t_{2n-1}$ is
performed with the help of \be
  \int \frac{d\tau_1}{\pi} e^{2i\Phi(\tau_1)}
  \frac{1}{\tau_0-\tau_1} \frac{1}{\tau_1-\tau_2}
  =
  - i \frac{e^{2i\Phi(\tau_0)}-e^{2i\Phi(\tau_2)}}{\tau_0-\tau_2}
  - \pi e^{2i\Phi(\tau_0)} \delta(\tau_0-\tau_2) ,
\ee
that is obtained by regularizing $1/t=[1/(t-i0)+1/(t+i0)]/2$
and making use of analyticity of $e^{2i\Phi(\tau)}$
in the upper half-plane.
As a result, the integral in Eq.~(\ref{S-imag}) is transformed to
\be
  \int_{-\infty}^\infty \prod_{k=1}^n
  \left(
    \frac{d\tau_{2k}}{\pi} \:
    e^{-2i\Phi(\tau_{2k})}
  \right) \:
    \prod_{k=1}^n
    \left[
      - i \frac{e^{2i\Phi(\tau_{2(k-1)})}-e^{2i\Phi(\tau_{2(k+1)})}}
        {\tau_{2(k-1)}-\tau_{2(k+1)}}
      - \pi e^{2i\Phi(\tau_{2k})} \delta({\tau_{2(k-1)}-\tau_{2(k+1)}})
    \right] ,
\ee
where $\tau_0 \equiv \tau_{2n}$ and $\tau_{2(n+1)} \equiv \tau_{2}$.
Expanding the last product and combining similar terms
we rewrite it as
\be
  (-1)^n \delta(0) + \sum_{m=1}^n C_n^m (-1)^{n-m} K_{m} ,
\label{binom}
\ee
where
\be
  K_m = (-i)^m
  \int_{-\infty}^\infty \prod_{k=1}^m
  \left(
    \frac{d\tau_{2k}}{\pi} \:
    e^{-2i\Phi(\tau_{2k})}
  \right) \:
    \prod_{k=1}^m
       \frac{e^{2i\Phi(\tau_{2(k-1)})}-e^{2i\Phi(\tau_{2(k+1)})}}
        {\tau_{2(k-1)}-\tau_{2(k+1)}} .
\ee
The integrals for $K_m$ are calculated recursively with the help
of the relation
\be
  \int \frac{d\tau_2}{\pi} e^{-2i\Phi(\tau_1)}
  \frac{e^{2i\Phi(\tau_1)}-e^{2i\Phi(\tau_2)}}{\tau_1-\tau_2}
  \frac{e^{2i\Phi(\tau_2)}-e^{2i\Phi(\tau_3)}}{\tau_2-\tau_3}
  =
  2 i
  \frac{e^{2i\Phi(\tau_1)}-e^{2i\Phi(\tau_3)}}{\tau_1-\tau_3} .
\ee
Thereby we get $K_m = 2^m W$. For anti--instantons with $W<0$
the same analysis yields $K_m = 2^m |W|$.

Now summation in Eq.~(\ref{binom}) becomes trivial
and we obtain for the instanton action:
\be
  S[\Phi(\tau)] =
  - \frac{G_D}{4} \sum_{n=1}^\infty \gamma_{2n}
  \Bigl\{
    (-1)^n \delta(0) + |W| [1-(-1)^n]
  \Bigr\} .
\label{S-inst}
\ee
Eq.~(\ref{S-inst}) formally contains the divergent
$\delta$-function of zero argument. This part of the answer gives
the action of the non-instanton configuration $\Phi=\Phi_0={\rm const}$
and thus drops from the difference ${\cal S}_W = S[\Phi(\tau)] - S[\Phi_0]$
which coincides with Eq.~(\ref{instaction}) obtained in the frequency domain.


\begin{thebibliography}{100}
\bibitem{actions}
The weight factor is $e^{iS[\Phi(t)]}$ in real time and
$e^{-S[\Phi(\tau)]}$ in imaginary time.
\end{thebibliography}
\end{document}